\documentclass[aps, prl, reprint, amsmath, longbibliography]{revtex4-1}
\usepackage{hyperref}
\usepackage{graphicx}
\usepackage[capitalize]{cleveref}
\usepackage{siunitx}
\AtBeginDocument{%
    \newwrite\bibnotes
    \def\bibnotesext{Notes.bib}
    \immediate\openout\bibnotes=\jobname\bibnotesext
    \immediate\write\bibnotes{@CONTROL{REVTEX41Control}}
    \immediate\write\bibnotes{@CONTROL{%
    apsrev41Control,author="08",editor="1",pages="1",title="0",year="1"}}
     \if@filesw
     \immediate\write\@auxout{\string\citation{apsrev41Control}}%
    \fi
}%

\begin{document}

\title{Spontaneous Crystallization in Systems of Binary Hard Sphere Colloids}

\author{Praveen K. Bommineni}
\author{Marco Klement}
\author{Michael Engel}\email{michael.engel@fau.de}
\affiliation{Institute for Multiscale Simulation, IZNF, Friedrich-Alexander University Erlangen-N\"urnberg, Cauerstrasse 3, 91058 Erlangen, Germany}

\date{\today}

\begin{abstract}
Computer simulations of the fluid-to-solid phase transition in the hard sphere system were instrumental for our understanding of crystallization processes.
But while colloid experiments and theory have been predicting the stability of several binary hard sphere crystals for many years, simulations were not successful to confirm this phenomenon.
Here, we report the growth of binary hard sphere crystals isostructural to Laves phases, AlB$_2$, and NaZn$_{13}$ in simulation directly from the fluid.
We analyze particle kinetics during Laves phase growth using event-driven molecular dynamics simulations with and without swap moves that speed up diffusion.
The crystallization process transitions from nucleation and growth to spinodal decomposition already deep within the fluid-solid coexistence regime.
Finally, we present packing fraction--size ratio state diagrams in the vicinity of the stability regions of three binary crystals.
\end{abstract} 
\maketitle


\emph{Introduction.---}Hard spheres are arguably one of the simplest particle model.
Their crystallization is a classic example of a phase transition discovered by computer simulation ~\cite{Alder1957a, Wood1957, Alder1960a}.
A prediction of the model is the spontaneous ordering of concentrated suspensions of nearly hard colloidal spheres~\cite{Pusey1986}.
Generalizations are mixtures of spheres with different sizes such that small spheres fit between large spheres to stabilize binary crystals.
Early experimental realizations are binary crystals isostructural to AlB$_2$ and NaZn$_{13}$ found in natural opal gems~\cite{Sanders1978,Sanders1980,Murray1980}.
More recent experiments~\cite{Bartlett1990,Bartlett1992,Hunt2000,Schofield2005,Schaertl2018}
and theory~\cite{Eldridge1993a,Eldridge1993,Eldridge1993b,Cottin1993,Cottin1995,TRIZAC1997,Hynninen2009,Dijkstra2014}
predicted additional ones.
By now four binary crystals have been proposed in the hard sphere phase diagram near the solidus line: NaCl ($0.30\leq\alpha\leq0.41$), AlB$_2$ ($0.45\leq\alpha \leq 0.62$), NaZn$_{13}$ ($0.54 \leq \alpha \leq 0.61$), and Laves MgCu$_2$ and MgZn$_2$ ($0.76 \leq \alpha \leq 0.84$).
This list identifies binary crystals by their atomic prototype and contains the reported ranges for size ratio $\alpha=\sigma_\text{S}/\sigma_\text{L}$, where $\sigma_\text{S}$ and $\sigma_\text{L}$ are the diameters of small and large spheres, respectively.
Additional binary crystals, such as CsCl~\cite{Schofield2005}, are believed to be metastable or appear only at high packing fraction $\phi$ as densest packings~\cite{RN5848, Filion2009, Hopkins2012}.

Equal-sized hard spheres crystallize rapidly into the face-centered-cubic (fcc) crystal or stacking variants thereof in simulation, and they have been an ideal testing ground for studying fundamental aspects of crystal nucleation and growth~\cite{Auer2001,Punnathanam2006,Ni2011}.
It was thus expected that simulations of binary hard sphere fluids produce binary crystals in a similar manner.
This has not been the case.
To date, the only binary hard sphere crystal growth reported in simulation is NaCl~\cite{TRIZAC1997}.
That report is more than 20 years old and the crystals grown are highly defective with many vacancies of the small spheres~\cite{Vermolen2009,Filion2011}.
AlB$_2$ so far required a seeding procedure to grow~\cite{Bommineni2017a}.
Laves phases and NaZn$_{13}$ formed from size-disperse sphere fluids only with the assistance of Monte Carlo swap or resize moves~\cite{Lindquist2018a,Bommineni2019} and nearly hard spheres where particle softness enhances crystallization~\cite{Dasgupta2019}.
The absence of binary hard sphere crystallization in simulation has been puzzling.
Here, we report the spontaneous formation of AlB$_2$, NaZn$_{13}$, and Laves phases in simulation directly from the fluid.
Our results demonstrate that binary hard sphere crystals grow robustly and reproducibly in standard event-driven molecular dynamics (EDMD) given only sufficient simulation time and large enough system size.
Surprisingly, the Laves phase crystallizes via nucleation and growth as well as via spinodal decomposition.
We find that the bottleneck for binary crystal growth is diffusion in the dense fluid.


\emph{Growth of Laves phase.---}Laves phases are of relevance for materials scientists because they are the most common binary intermetallic compounds~\cite{Steurer2016} and have interesting photonic properties when self-assembled from colloids~\cite{Hynninen2007}.
For this reason, we focus on their growth first.
We speed up crystallization in an initial test by combining EDMD simulations with swap moves~\cite{Berthier2016,Wyart2017,Lindquist2018a,Coslovich2018,Brito2018,Bommineni2019,Berthier2019,Mihalkovic2019}.
Particle pairs are attempted to be swapped as a Monte Carlo move at every collision~\cite{SuppMat}.
Gibbs free-energy calculations~\cite{Hynninen2009} guide us to the parameter set $(\phi,\alpha)=(0.57,0.80)$, which lies in the fluid-Laves coexistence regime slightly below the solidus line.
We initialize a simulation at composition LS$_2$ in the fluid state and run it at isochoric (constant volume) conditions with periodic boundaries.
The ordering progress of our system is monitored by recording reduced pressure $P^\ast= P / \Pi$ over reduced simulation time $t^\ast=t / \tau$.
Here, $\Pi = k_\text{B} T / V_\text{L}$ with volume of the large sphere $V_\text{L}=\pi \sigma_\text{L}^3/6$ is a unit of pressure and $\tau=\sigma_\text{L} \sqrt{m/{k_\text{B}T}}$ is a unit of time.

\begin{figure*}
\includegraphics[width=\linewidth]{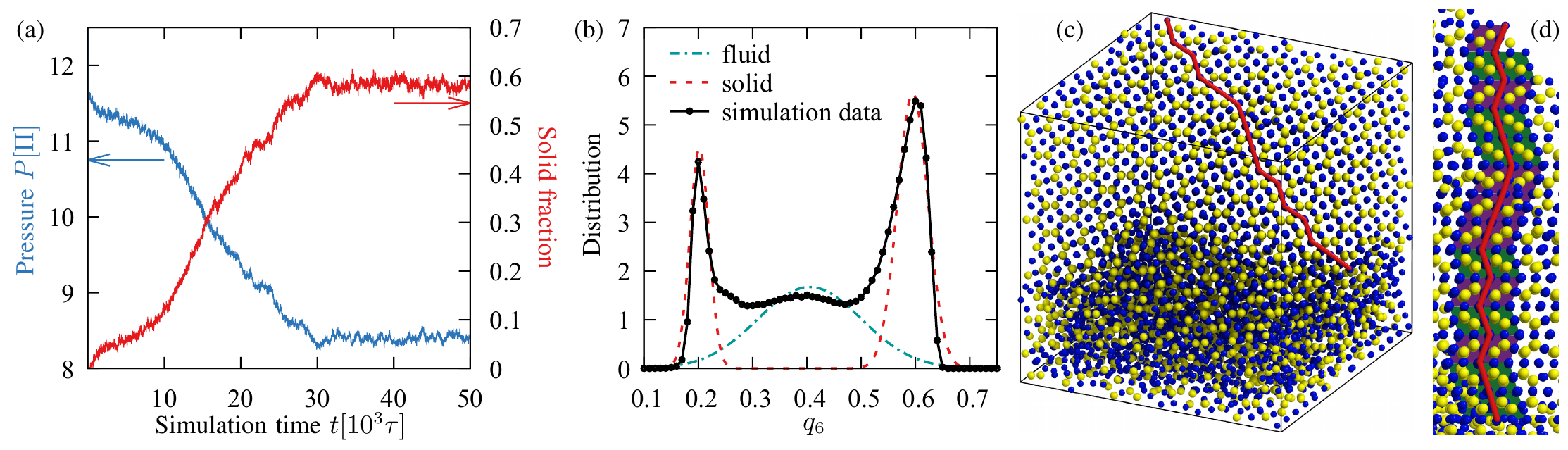}
\caption{Crystallization of the Laves phase from a binary hard sphere fluid using a hybrid EDMD simulation with swap moves.
The system contains 9999 particles at composition LS$_2$ and is simulated at $(\phi,\alpha)=(0.57,0.80)$.
(a)~Evolution of pressure (blue) and solid fraction (red) with simulation time. 
(b)~Distribution of local bond-orientational order parameters $q_6$ at fluid-solid coexistence.
Solid fraction is computed by fitting the distribution with three Gaussians as indicated by dashed curves.
(c)~Final snapshot and (d)~enlarged view of the simulation showing the coexistence of Laves phase and fluid.
Large spheres are represented by yellow color, small spheres by blue color.
Particles are drawn at 50\% of their size for better visibility.
Red line indicates the stacking sequence.}
\label{fig:Pressure}
\end{figure*}

Pressure evolution is shown as blue curve in \cref{fig:Pressure}(a).
After a rapid initial relaxation of the fluid over time $3\times10^3\tau$, pressure decreases slowly while the system starts ordering, then gradually faster, until it reaches a constant slope.
Crystalline order in the system is well characterized by the $q_6$ local bond-orientational order parameter~\cite{Steinhardt1983,vanMeel2012,SuppMat}.
The distribution of $q_6$ values in the system shows pronounced peaks (\cref{fig:Pressure}(b)), which can be assigned to the fluid (broad central peak) and the two particle species in the solid (narrow outer peaks).
We fit the peaks to compute solid fraction.
Evolution of solid fraction (red curve in \cref{fig:Pressure}(a)) essentially mirrors the evolution of pressure and increases in sync with the pressure drop.
The slopes of both curves come to an abrupt halt at $t=30\times10^3\tau$, indicating the end of the crystallization process.
After this time, the system reached an equilibrium of about equal amounts of fluid and solid in coexistence.

A snapshot of the equilibrated system viewed along a twofold symmetry axis (\cref{fig:Pressure}(c)) confirms the coexistence.
In projection, particles are arranged into columns in the crystal (top part) and are disordered in the fluid (bottom part).
The pattern of the crystal in this projection consists of straight rows of small spheres separated by zigzag rows of pairs of large spheres alternating with single small spheres.
Such a pattern is characteristic of Laves phases.
We extract the stacking sequence of the Laves phase as indicated by the red line in \cref{fig:Pressure}(c,d).
Hexagonal Laves MgZn$_2$ (C14) has ABA stacking, which gives a zigzag line.
Cubic Laves MgCu$_2$ (C15) on the other hand has ABC stacking, which results in a straight line.
We observe both straight and zigzag segments along the red line, demonstrating that our Laves phase is a stacking of C14 slabs and C15 slabs.
In this sense, Laves phase crystallization in our binary system resembles crystallization of identical spheres, which forms stacking variants of fcc.

\begin{figure*}
\includegraphics[width=0.95\linewidth]{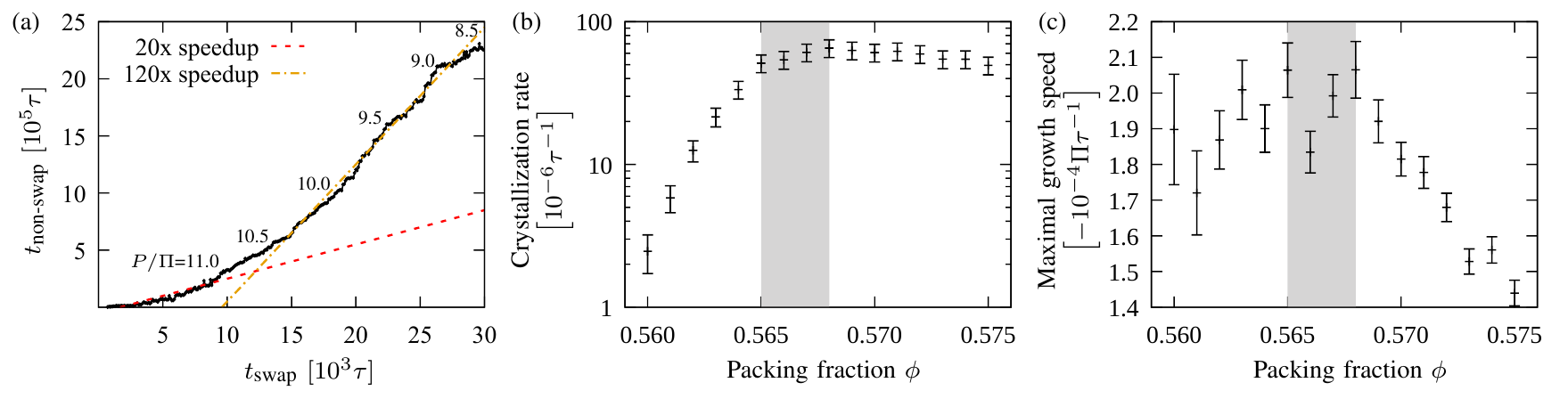}
\caption{Laves phase growth kinetics.
Simulation parameters and conditions as in \cref{fig:Pressure}.
(a)~Simulation times necessary to reach certain pressure values in swap and nonswap simulations are compared on the two plot axes.
The speed-ups for early and late crystallization are indicated by dashed lines.
(b)~Crystallization rate gives the inverse mean time to find a growing crystal in a simulation.
As density increases, we see a transition from rare nucleation events to nearly constant crystallization rates for spinodal decomposition.
Error bars indicate the spread assuming nucleation and growth as the mechanism.
(c)~Maximal growth speed is defined as the fastest pressure drop and is highest in the transition region (gray area).
Weak oversaturation lowers the entropy gain of an ordered structure, and high density slows diffusion.
Both factors are balanced at the onset of spinodal decomposition.
Data points in (a) and (b,c) are averaged over 10 and 50 runs, respectively.
Error bars show the standard error.}
\label{fig:LavesGrowthKinetics}
\end{figure*}

\emph{Laves phase crystallization pathway.---}Our simulation in \cref{fig:Pressure} established Laves phase growth from the binary hard sphere fluid using swap moves rather easily.
To show reproducibility and generality, we perform ten swap simulations and ten nonswap simulations at the same parameters and with similar initial conditions.
Well-ordered Laves phases form in all 20 simulations.
Both simulation methods lead to similar pressure-time pathways (Fig.~S1), even though the speed in which these pathways are traversed is different.
As expected, swap simulations are significantly faster in crystallizing the binary fluid.
The speed-up is not constant but increases from $20\times$ to $120\times$ (\cref{fig:LavesGrowthKinetics}(a)), demonstrating that the efficiency of swap moves improves over time.
Indeed, the acceptance probability of swap moves increases over the same time window (Fig.~S2).
This indicates an increase in available free volume in the fluid and explains the higher speed=up toward the end of the simulation.
Swap moves primarily enhance diffusion.
We conclude that particles are integrated faster into the crystal than they diffuse through the fluid.
Therefore, diffusion in the fluid is the bottleneck process for Laves phase growth.

\begin{figure}
\centering
\includegraphics[width=0.86\linewidth]{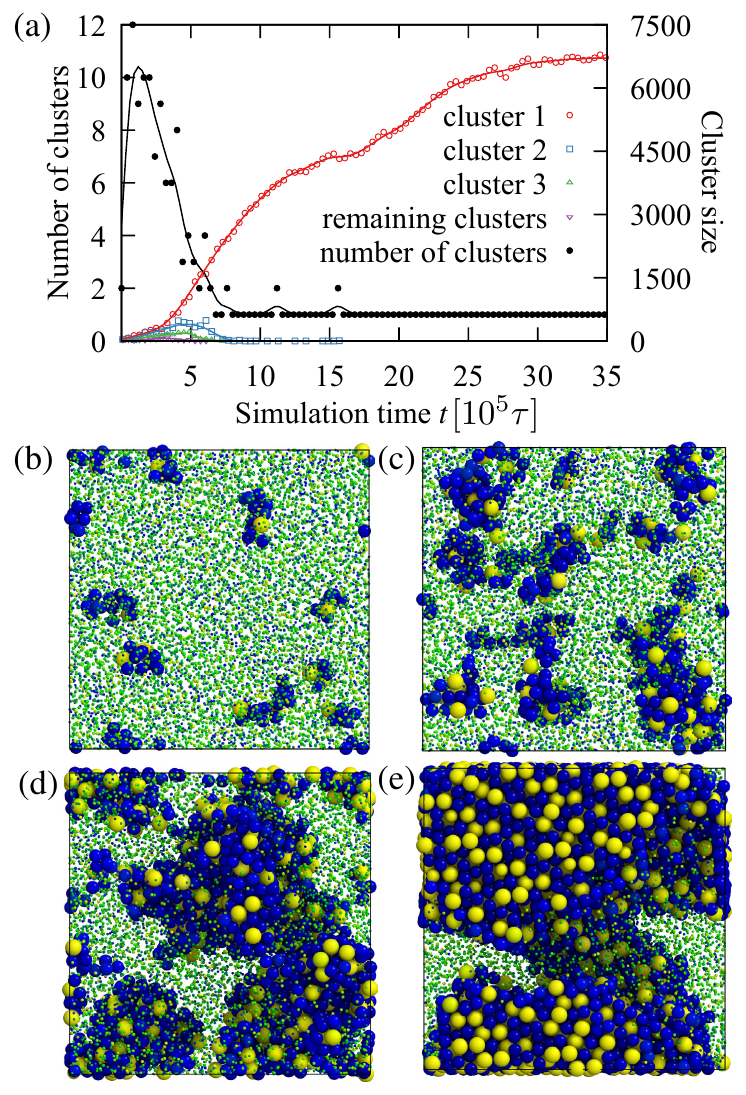}
\caption{Exemplary Laves phase growth pathway from a nonswap simulation.
Simulation parameters and conditions as in \cref{fig:Pressure}.
(a)~Evolution of the number of clusters and the sizes of the three largest clusters ('cluster 1' to 'cluster 3').
Simulation snapshots at times (b)~$t=2\times10^3\tau$, (c)~$t=10^5\tau$, (d)~$5\times10^5\tau$, and (e)~$2\times10^6\tau$.
Particles are colored according to their local environment.
Blue particles have high $q_6$, yellow particles low $q_6$.
Green particles have intermediate $q_6$ corresponding to a fluid-like environment.
Nonclustered particles are drawn with reduced size for better visibility.
}
\label{fig:Growth}
\end{figure}

We analyze one exemplary nonswap simulation pathway in more detail by tracking crystalline clusters.
\cref{fig:Growth}(a) shows the evolution of the number of clusters and the size of the three largest clusters.
A particle is identified as solid-like using a $q_6$ cut-off (Fig.~S3).
A solid-like particle belongs to a cluster if it has more than five solid-like particles within distance $1.1\sigma_\text{L}$.
Already right at the start of the simulation, ten clusters are detected, indicating there is a very small or negligible free-energy barrier for the Laves phase to form. 
The clusters grow independently, and their number decreases when they merge, which is also directly apparent in simulation snapshots at increasing times (\cref{fig:Growth}(b-e)).
After this time, the solid consolidates by removing grain boundaries and transforms its polycrystalline state into a Laves phase single crystal.
A video of the growth process for the time window $0\le t\le 2\times 10^6\tau$ is contained as Supplemental Material.
Overall, Laves phase growth in this simulation proceeds not as predicted by classical nucleation theory but as expected from spinodal decomposition.
This is surprising given that our simulation parameters are chosen below the solidus line in the coexistence regime (Fig.~S4).

Our results so far do not exclude a nucleation and growth regime at lower packing fraction closer to the liquidus line.
We analyze the type of crystallization transition by performing $15\times50$ swap simulations across the packing fraction range $0.56\le\phi\le0.575$.
Indeed, the crystallization rate grows rapidly at $\phi\le 0.565$ as expected for nucleation and growth and then saturates (\cref{fig:LavesGrowthKinetics}(b)).
At the same time, the maximal crystal growth speed decays from a plateau for $\phi\ge 0.568$ as expected for spinodal decomposition (\cref{fig:LavesGrowthKinetics}(c)).
Together, our analysis indicates a transition from nucleation-and-growth behavior with stochastic onset of crystallization to spinodal-decomposition behavior where crystallization pathways collapse.
This transition is also apparent directly in the evolution of pressure curves (Fig.~S5).

\begin{figure*}
\centering
\includegraphics[width=1\linewidth]{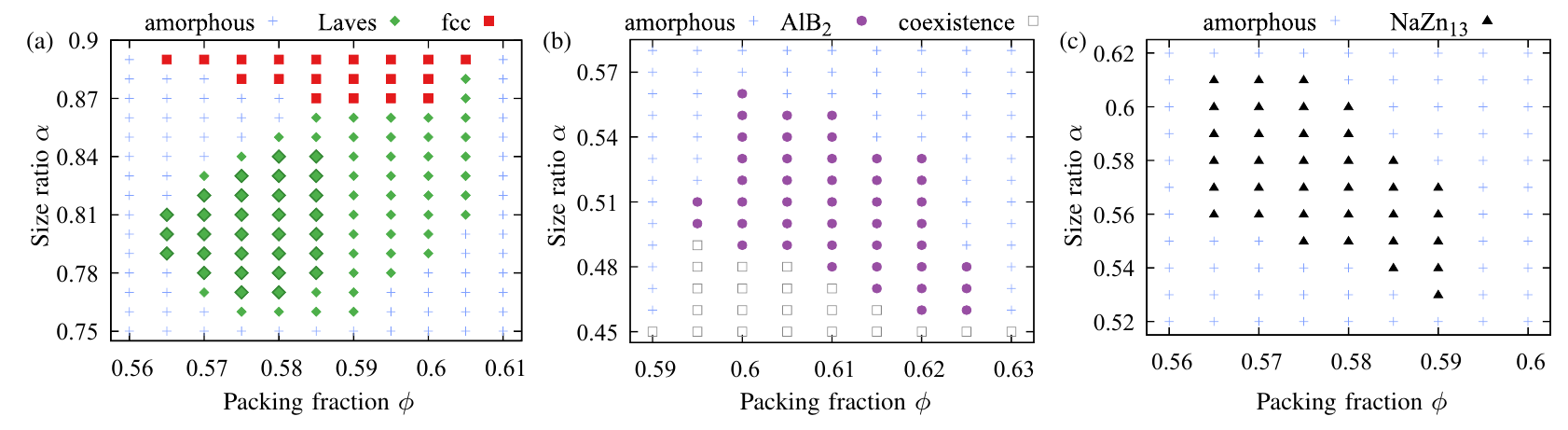} 
\caption{Packing fraction--size ratio state diagrams in the vicinity of the stability regions of the three binary hard sphere crystals isostructural to (a)~Laves phases, (b)~AlB$_2$, and (c)~NaZn$_{13}$.
Crystal structures are identified using bond-orientational order diagrams and directly in snapshots (Fig.~S6, S7,~\cite{Engel2015}).
Systems contain 1200 particles at composition LS$_2$ in (a,b) and 896 particles at composition LS$_{13}$ in (c).
Simulations in (a) are run with swaps, large symbols represent Laves phase formation in nonswap simulations.
Nonswap simulations crystallize less readily, which is apparent near the boundary of the stability regime and at high packing fraction.
Simulations in (b,c) are run without swaps because the acceptance probability of swap moves is too low for swaps to be useful.
At each state point, simulations are repeated twice, and phases reported are observed at least once in simulation. 
}
\label{fig:Statediagrams}
\end{figure*}

\emph{Binary crystal state diagrams.---}So far, we investigated binary crystal growth for a specific set of parameters in detail.
Now, we vary the packing fraction $\phi$ and size ratio $\alpha$ more systematically to obtain full state diagrams.
We focus on the compositions LS$_2$ and LS$_{13}$, for which binary crystals have been predicted.
Simulations run for a total time of $2\times10^6\tau$.
Simulations that crystallize undergo their phase transition completely and fully grow into well-ordered crystals.

We start with composition LS$_2$.
\cref{fig:Statediagrams}(a) shows the state diagram in the vicinity of Laves phases and \cref{fig:Statediagrams}(b) the state diagram in the vicinity of AlB$_2$.
Both binary crystals form over extended parameter regions.
The size ratio range for the Laves phases, $0.76 \leq \alpha \leq 0.86$, and the lower limit of size ratio for AlB$_2$, $\alpha=0.46$, agree with predicted values~\cite{Hynninen2009,Eldridge1993a}.
For $0.57\le\alpha\le0.75$, the system remains amorphous.
For $\alpha\le0.45$, the binary fluid phase separates into coexisting NaCl and AlB$_2$ solids.
For $\alpha\ge0.87$, the fluid forms the substitutionally disordered fcc solid.

Results for composition LS$_{13}$ are shown in \cref{fig:Statediagrams}(c). 
The binary crystal found at this composition, NaZn$_{13}$, is a complex arrangement of large spheres on a simple cubic lattice and icosahedra made from 13 small spheres at body-centered positions.
NaZn$_{13}$ forms in the range $0.54 \leq \alpha \leq 0.61$ as predicted by free-energy calculation~\cite{Eldridge1993b,Eldridge1993}.
The range of packing fractions, for which crystallization is observed, is for all three binary crystals about 3\%.
Growth of AlB$_2$ is more difficult and leads to more defects than growth of Laves phases and NaZn$_{13}$, possibly because it occurs at higher packing fractions.


\emph{Discussion and conclusions.---}Our findings demonstrate that binary crystal growth from the binary hard sphere fluid is robust and reproducible.
Why then have similar binary hard sphere crystals not been seen in simulation before?
We are not sure.
But we can list a few arguments that indicate doing so is not necessarily trivial.
In simulations, growth without swap moves requires more than two weeks of continuous simulation with today's fastest processors at the right parameters to obtain well-defined crystals that are easy to identify.
Our longest simulations ran for more than four weeks.
While this is not a particularly long simulation time, it clearly exceeds typical hard sphere simulation times.
EDMD simulations cannot be parallelized efficiently.
Furthermore, many simulations in the past used Monte Carlo simulation instead of EDMD, which slows down the growth process further by about one order of magnitude~\cite{Klement2019}.

We map our simulations on two experimental conditions: magnesium atoms and polystyrene colloids.
The unit of time for magnesium atoms at a temperature slightly below the solidus line for hexagonal Laves MgZn$_2$, $T=\SI{850}{\kelvin}$, corresponds to $\tau=\SI{0.6}{\pico\second}$~\cite{SuppMat}.
An atom takes on average about the time $\tau$ to move over its diameter.
We estimate that Laves phase crystallization requires an experimental time of $10^7\tau=\SI{10}{\micro\second}$ to form nanocrystalline grains.
Such times are easily reachable.
This explains why Laves phases are ubiquitous in alloys.
Colloidal particles on the other hand are much larger and slowed down by drag in solution.
It is possible to account for most of the effects of hydrodynamic interactions by rescaling the timescale for nucleation and growth by the long-time diffusion coefficient~\cite{Tateno2019}.
We use the Stokes-Einstein equation to estimate $\tau=\SI{0.2}{\second}$ for colloids with diameter $\SI{1}{\micro\meter}$ suspended in water~\cite{SuppMat}.
Only crystallization of identical hard sphere colloids into fcc has been achieved with such large particles as it is about $10^3$ times faster than Laves phase crystallization, even though the softness of charged spheres seems to help obtain a variety of binary structures~\cite{Hachisu1980,Bartlett2005,Leunissen2005,LaCour2019}.
Crystallization speeds up proportionally to $\sigma^{-3}$ with shrinking colloid diameter $\sigma$.
Experiments with $\sigma_\text{L}=\SI{170}{\nano\meter}$ colloids need to rest for months at the optimal condition to form Laves phases~\cite{Schaertl2018}, which corresponds to an equilibration time of $10^9\tau$ to $10^{10}\tau$ to grow macroscale crystals.
Even smaller $\le\SI{10}{\nano\meter}$ nanoparticles crystallize rather easily into binary superlattices~\cite{Shevchenko2005,Shevchenko2006,Evers2010a,Coropceanu2019}.

There have been concerns that swaps alter phase transformation pathways because they preferentially accelerate specific aspects of kinetics.
At least for the processes investigated here, crystallization of hard sphere Laves phases, trajectories with and without swap are indistinguishable.
If these observations are confirmed in more systems, then swaps~\cite{Berthier2016,Wyart2017,Lindquist2018a,Coslovich2018,Brito2018,Bommineni2019,Berthier2019,Mihalkovic2019} can become a standard tool for simulating mixtures, in particular those with small size difference.
Swap methods have the potential to speed up such simulations by orders of magnitude.
Finally, our findings add to the growing list of recent observations of crystallization processes in model systems that were once believed to be good glass formers~\cite{Pedersen2018,Coslovich2018,Bommineni2019,Ingebrigtsen2019,Dasgupta2019}.
Knowledge of crystal structures competing with the amorphous state, as we report here for the binary hard sphere fluid, is important to ensure that the analysis of local order and particle dynamics~\cite{Robinson2019,Marin-Aguilar2019,Campo2019} in such model systems intended for glasses is not affected by hidden crystallization transitions.

In conclusion, this work constitutes the first systematic simulation study of binary crystallization across composition and thermodynamic parameters.
Our finding of a transition from nucleation and growth to spinodal decomposition inside the fluid-solid coexistence demonstrates a clear difference of the crystallization behavior of binary hard sphere mixtures as compared to the crystallization behavior of identical hard spheres.

\begin{acknowledgments}
Funding by Deutsche Forschungsgemeinschaft through Project No. EN 905/2-1, support from the Central Institute for Scientific Computing, the Interdisciplinary Center for Functional Particle Systems, and computation resources provided by the Erlangen Regional Computing Center are gratefully acknowledged.
\end{acknowledgments}

\renewcommand{\thefigure}{S\arabic{figure}}
\setcounter{figure}{0}    


\begin{figure}
\includegraphics[width=\linewidth]{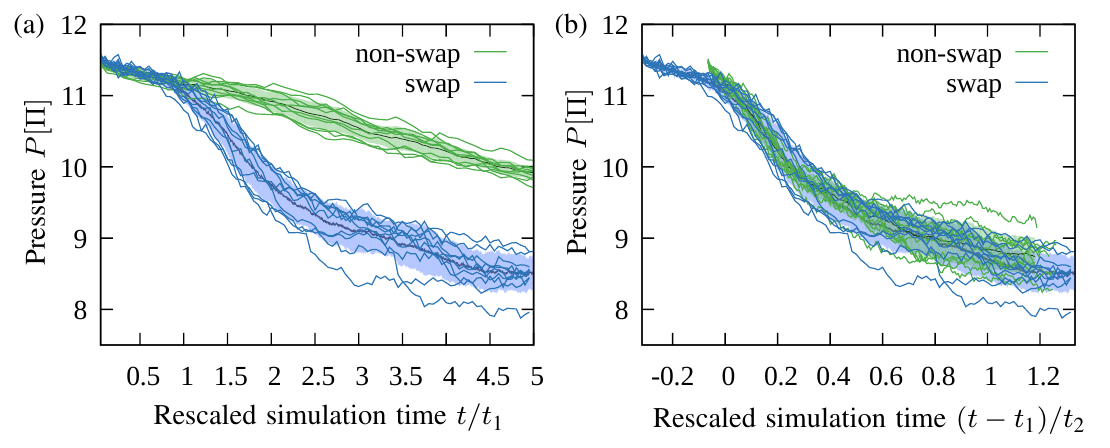}
\caption{Comparison of Laves phase growth pathways from ten swap simulations and ten non-swap simulations.
(a)~Initial pressure evolution is matched by rescaling simulation times by $t_1 = 10^4\tau=2.3\text{ CPU hours}$ for swap simulations and $t_1 = 20\times10^4\tau = 45\text{ CPU hours}$ for non-swap simulations.
Note that the tails of the pathway do not match.
(b)~Tails of pressure evolution are matched by rescaling simulation times by $t_2 = 3\times10^4\tau$ for swap simulations and $t_2 = 250\times10^4\tau$ for non-swap simulations.
Note that the initial parts do not match.
Simulation parameters and conditions as in Fig.~1.
Data in this figure forms the basis of Fig.~2a.}
\label{fig:LavesGrowthPathways}
\end{figure}

\begin{figure}
\includegraphics[width=0.83\linewidth]{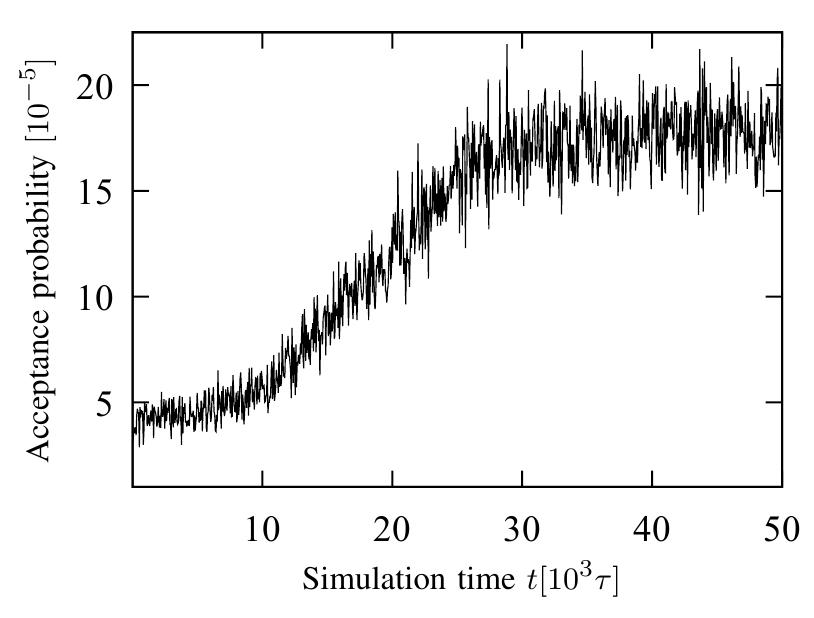}
\caption{Evolution of particle swap acceptance probability for the trajectory shown in Fig.~1.
Particle swap acceptance probability is defined as the number of accepted particle swaps to the number of attempted particle swaps. 
Acceptance probability gradually increases from $5\times10^{-5}$ to $20\times10^{-5}$ while the system crystallizes.}
\label{fig:swapAcceptance}
\end{figure}

\begin{figure}
\includegraphics[width=0.75\linewidth]{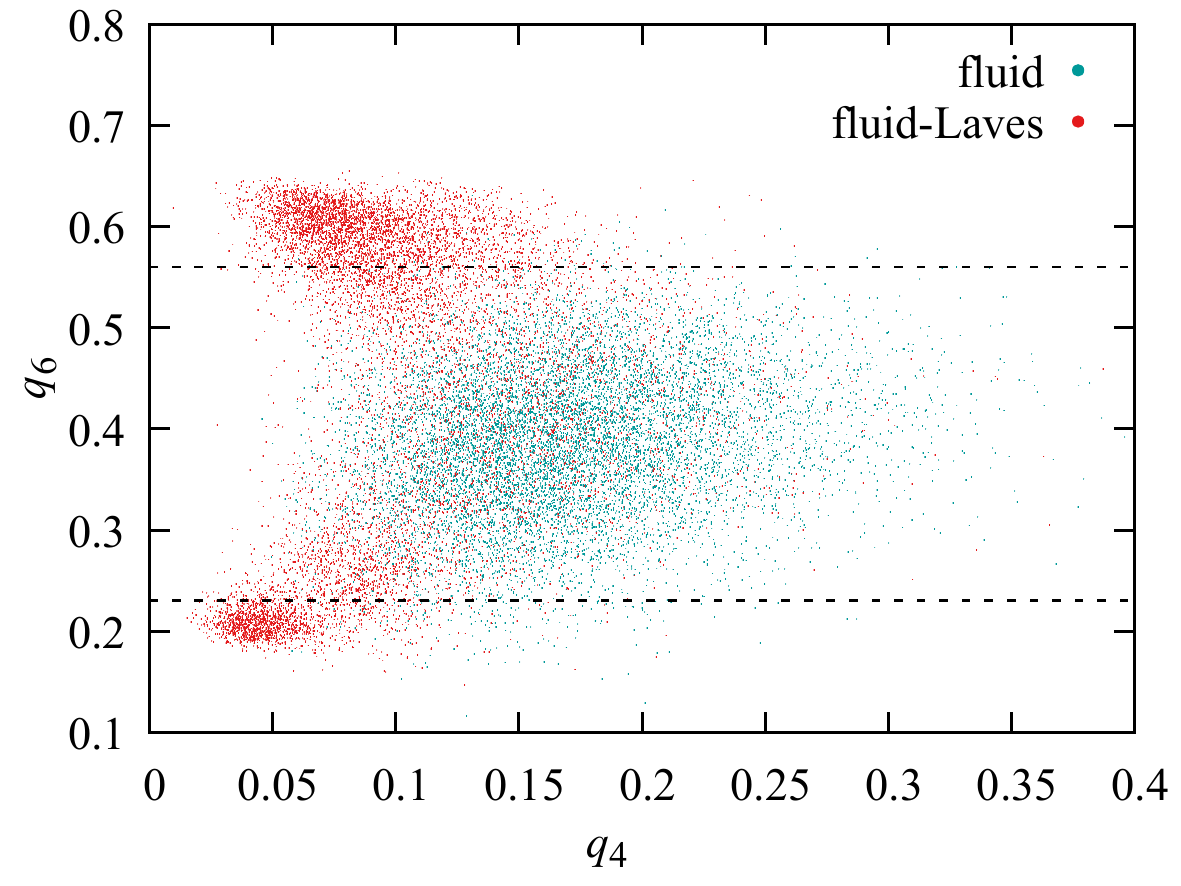}
\caption{Scatter plot of $q_4$ and $q_6$ local bond-orientational order parameter values of all 9999 particles in a fluid configuration (cyan) and in a coexisting fluid-Laves configuration (red).
Both configurations are taken from the trajectory in Fig.~1. 
Dashed lines indicate $q_6$ cut-off used to identify solid-like particles.}
\label{fig:q_scatter}
\end{figure}

\begin{figure}
\includegraphics[width=0.7\linewidth]{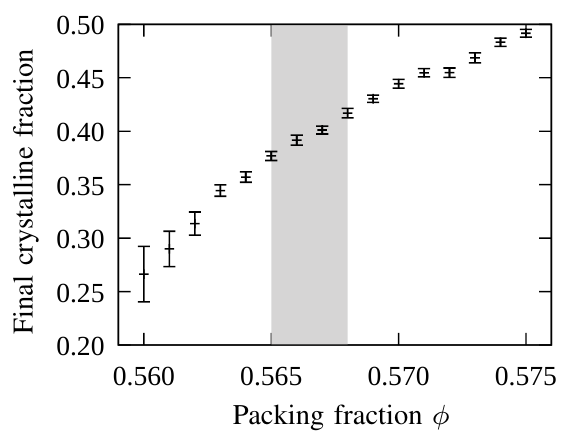}
\caption{Fraction of particles in the Laves phase crystallization simulations of Fig.~2(b,c).
The figure shows the number of crystalline particles at the end of crystallization simulations averaged over 50 simulation runs.
It is apparent that the transition from nucleation-and-growth behavior to spinodal-decomposition behavior, indicated by a gray area, occurs deep within the fluid-solid coexistence.}
\label{fig:FinalCrystallineFraction}
\end{figure}

\section{Appendix A: Simulation method}
Binary hard spheres are simulated in event-driven molecular dynamics (EDMD) with and without swap as described in recent work~\cite{Bommineni2019}.
Setting the mass equal for all $N$ particles simplifies momentum conservation and has no effect on the phase behavior of the system.
Dimensionless pressure $P^\ast$ is computed as~\cite{Alder1960a}
\begin{equation}
	P^\ast = \frac{P \pi \sigma_\text{L}^3 }{6 k_B T} = \phi\left(1 + \frac{\sqrt{\pi}}{3} \frac{\tau}{t_\text{tot}} \frac{N_\text{c}}{N}\right),
\end{equation}
from the number of particle collisions $N_\text{c}$ during a simulation time window $t_\text{tot}$.
$k_\text{B}T$ is the product of the Boltzmann constant $k_\text{B}$ and the temperature $T$.

\begin{figure}
\includegraphics[width=\linewidth]{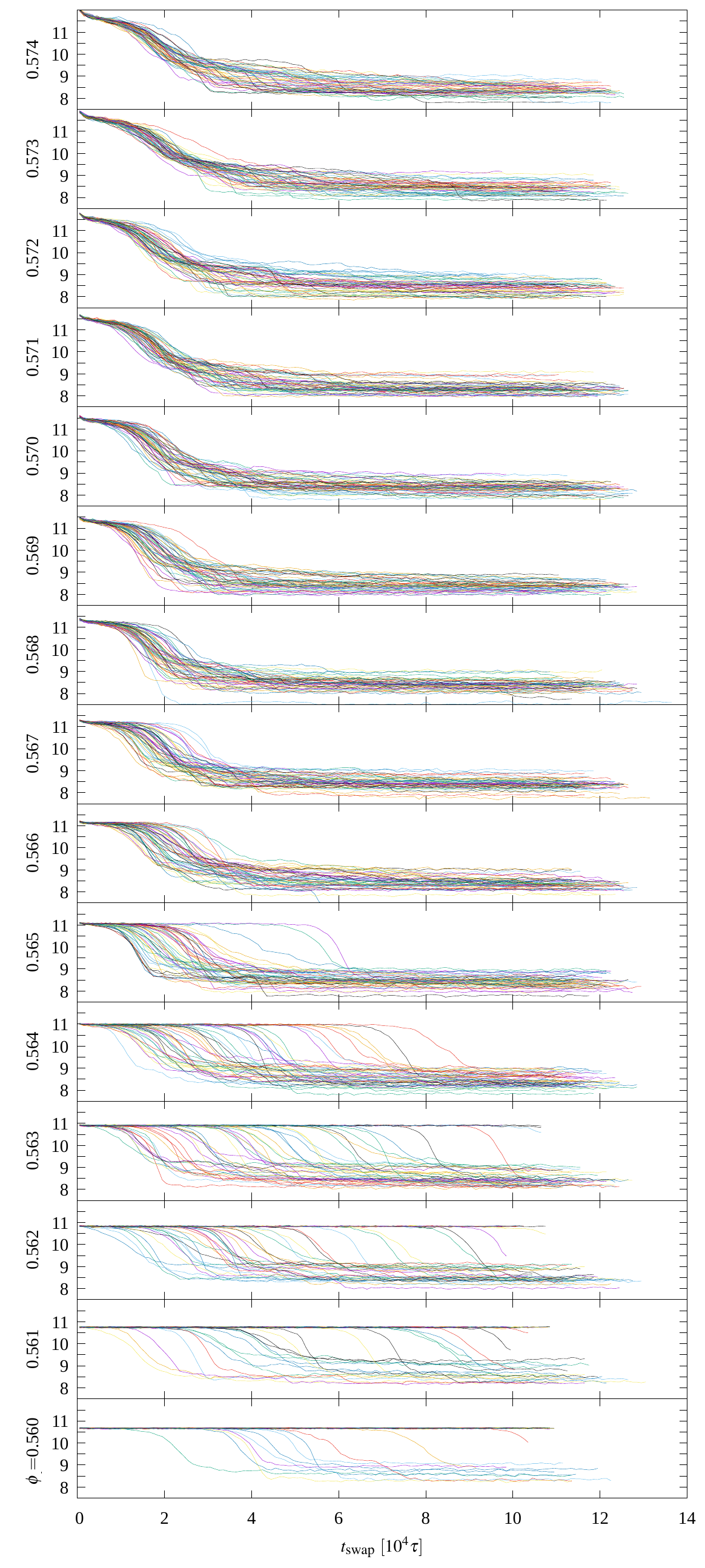}
\caption{Evolution of dimensionless pressure $P^\ast$ ($y$-axes) vs. simulation time $t_\text{swap}$ ($x$-axis) for the $15\times 50$ swap simulations of Fig.~2(b,c).
Delayed crystallization in the form of plateaus with stochastic nucleation events is apparent in the packing fraction range $\phi\le0.565$.
At higher packing fractions, at $\phi\ge 0.568$, pressure curves collapse as expected for spinodal decomposition.
For spinodal decomposition there is an initial, deterministic delay of crystallization, which indicates that the fluid needs to prestructure to some degree before ordering can proceed more rapidly.
Note that the plateau is never fully flat and crystallization curves always collapse for spinodal decomposition.
Pressure data in this figure was averaged over $10^3\tau$.}
\label{fig:PressureEvolution}
\end{figure}

\begin{figure}
\includegraphics[width=0.7\linewidth]{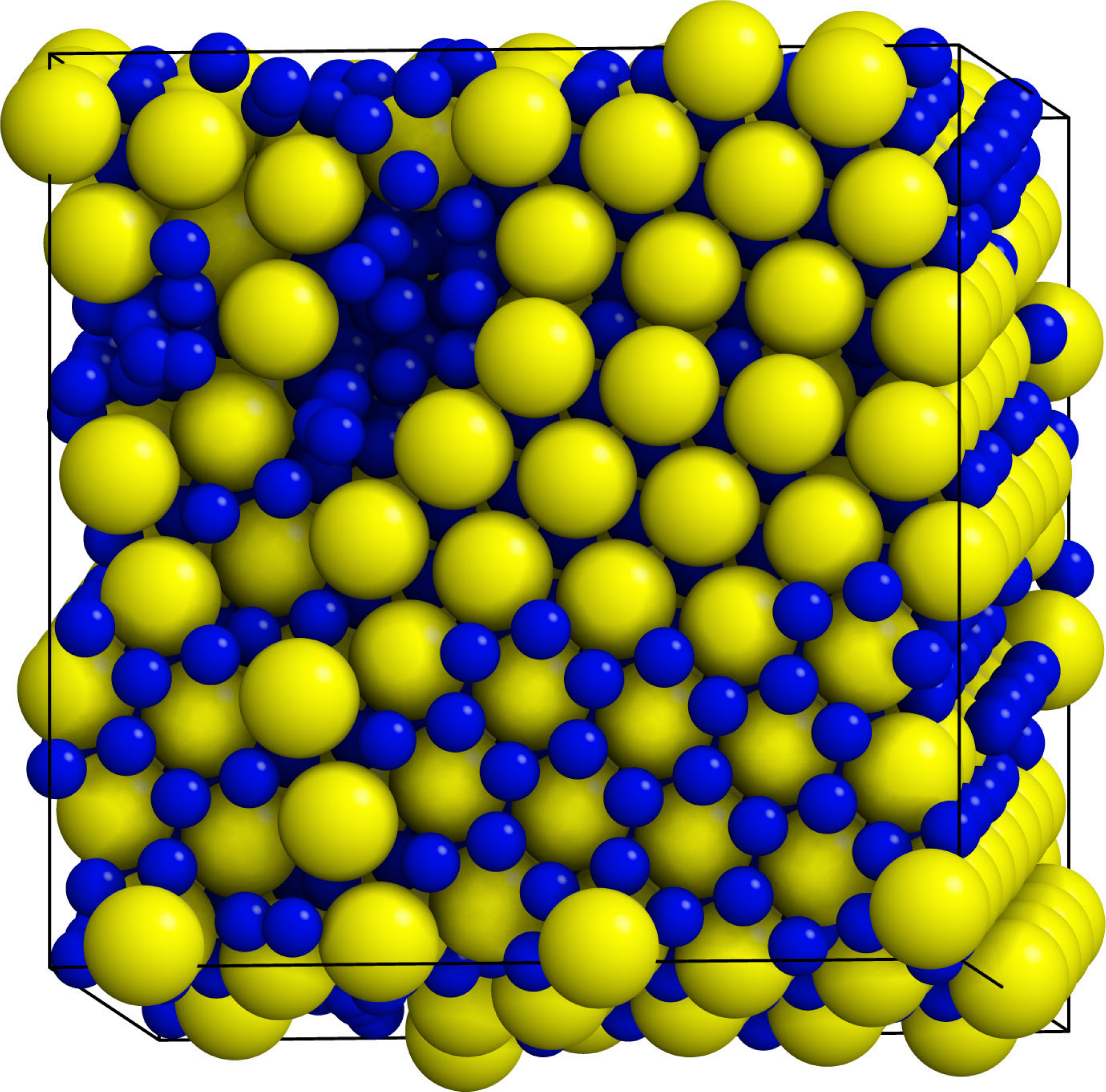}
\caption{Snapshot of a binary crystal isostructural to AlB$_2$ formed from hard sphere simulations at $(\phi,\alpha)=(0.60,0.50)$.
AlB$_2$ consists of hexagonal rings of small particles (blue) filled by large particles (yellow).}
\label{fig:AlB2}
\end{figure}

\begin{figure}
\includegraphics[width=0.7\linewidth]{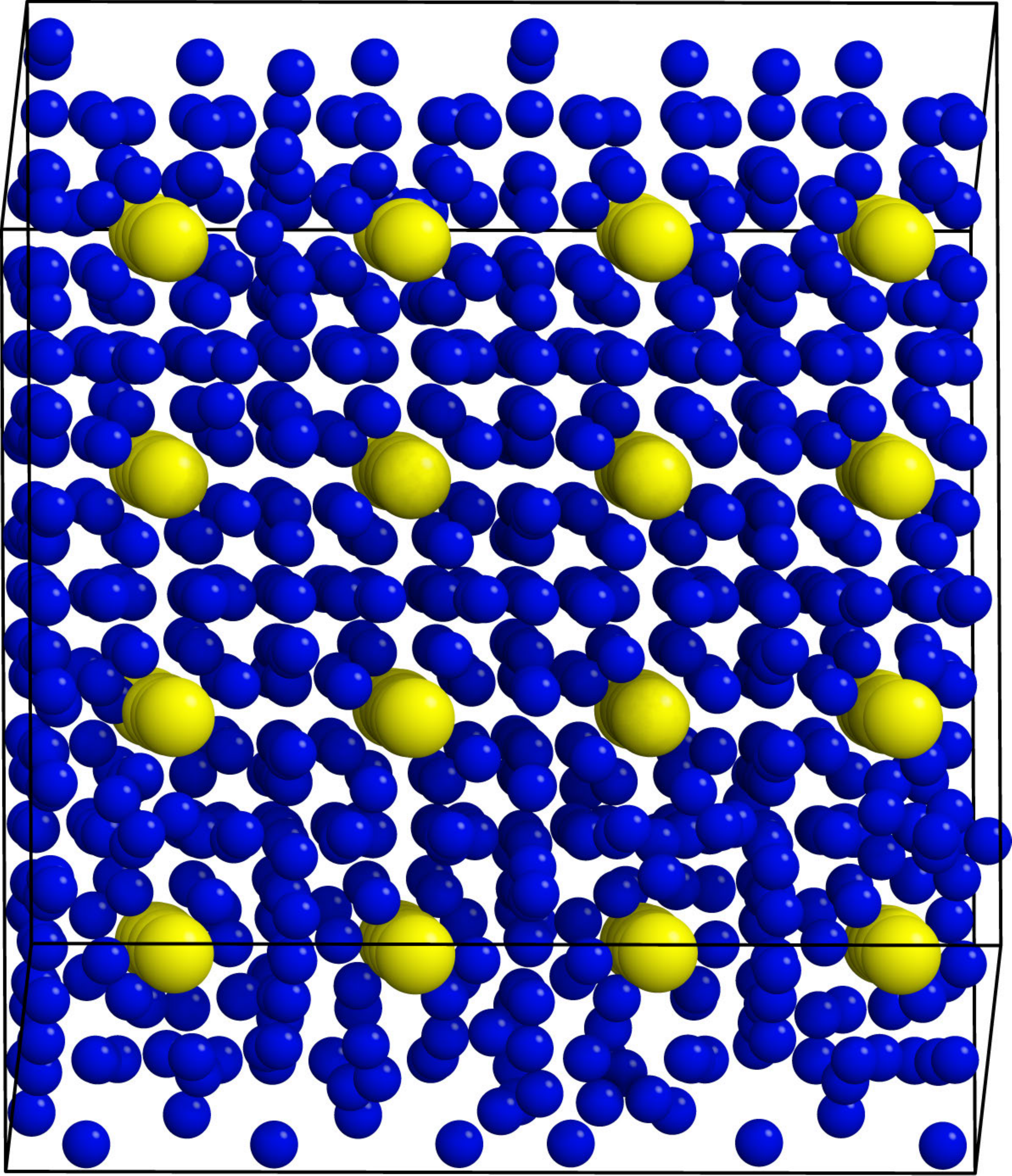}
\caption{Snapshot of a binary crystal isostructural to NaZn$_{13}$ formed from simulations at $(\phi,\alpha)=(0.57,0.60)$. Large particles arranged on simple cubic unit cell and 13 small particles with icosahedral symmetry at body-centered positions.
Particles are drawn at 50\% of their size for better visibility.}
\label{fig:AB13}
\end{figure}

\section{Appendix B: Detection of crystalline clusters}
Crystalline clusters growing from the binary hard sphere fluid are tracked with local bond-orientational order parameters $q_l$~\cite{Steinhardt1983}.
We compute the $(2l+1)$-dimensional complex vectors
\begin{equation}
q_{l}^{m}(i) = \frac{1}{N_b(i)} \sum_{j=1}^{N_b(i)} Y_{l}^{m}\left(\vec{r}_{ij}\right).
\end{equation}
Here, $N_b(i)$ is the number of neighbors of particle $i$, $Y_{lm}(\vec{r}_{ij})$ are spherical harmonics coefficients for neighboring particles $j$, and $\vec{r}_{ij}$ is the separation vector. 
Nearest neighbor particles are identified using the parameter free solid-angle based nearest-neighbor (SANN) algorithm~\cite{vanMeel2012}.
Local order parameters are defined as
\begin{equation}
q_{l}(i) = \sqrt{\frac{4 \pi}{2l+1} \sum_{m=-l}^{l} {|q_{l}^{m}(i)|}^2}.
\end{equation}
Fig.~S2 shows the scatter plot of $q_4$ and $q_6$ for all particles in a fluid and in a fluid-solid coexistence of Fig.~1. 
$q_6$ alone is sufficient to detect whether a small or large particle is in a fluid-like environment or a solid-like environment. 
Particles are identified as solid-like if $q_6 \leq 0.23$ or $q_6 \geq 0.56$ (indicated by dashed line in Fig.~S2).
Finally, clusters are extracted by grouping solid-like particle with more than five solid-like neighboring particles within a cut-off distance $1.1\sigma_\text{L}$.

\section{Appendix C: Mapping simulations to experiments}
We perform a highly simplistic mapping of our simulation trajectories to two experimental situations, magnesium atoms and polystyrene colloids in water.
For magnesium atoms, we assume ballistic motion between collisions and use the van der Waals radius $\sigma_\text{L}/2=\SI{173}{\pico\meter}$ and mass $m=\SI{24.3}{\dalton}$ of a magnesium atom and a temperature slightly below the solidus line for hexagonal Laves MgZn$_2$, $T=\SI{850}{\kelvin}$, to calculate
\begin{eqnarray}
\tau&=&\sigma_\text{L} \sqrt{\frac{m}{k_\text{B}T}}=\SI{0.6}{\pico\second},\\
\Pi&=&\frac{6k_\text{B} T} {\pi \sigma_\text{L}^3} = \SI{5}{\kilo\bar}.
\end{eqnarray}
Furthermore, from the Maxwell-Boltzmann distribution we know that the average atom velocity is
\begin{equation}
\langle v\rangle = \sqrt{\frac{8 k_\text{B} T} {\pi m}} = \sqrt{\frac{8}{\pi}}\frac{\sigma_\text{L}}{\tau}.
\end{equation}
In other words, the atom takes on average about the time $\tau$ to move over its diameter under the assumption of ballistic motion.

For polystyrene colloids with diameter $\sigma_\text{L}=\SI{1}{\micro\meter}$ in water, we use room temperature $T=\SI{293}{\kelvin}$ to calculate as before $\Pi = \SI{8}{\milli\pascal}$.
The assumption of ballistic motion is clearly not valid for colloidal particles in solution.
To compensate for diffusion in solution, we use the relation for the mean squared displacement and the Stokes-Einstein equation for the diffusion coefficient,
\begin{equation}
\frac{\langle x^2\rangle}{2t}=D=\frac{k_\text{B}T}{6\pi\eta \sigma_\text{L}}.
\end{equation}
We assume that particles move diffusively between collisions and average the slowdown from diffusion over all collisions $\text{c}$.
This means time is replaced by the average time between two successive collisions, $t=\langle\Delta t_\text{c}\rangle_\text{c}$, and mean squared displacement is replaced by the average squared distance the particle moves between two successive collisions, $\langle x^2\rangle=\langle\Delta x_\text{c}^2\rangle_\text{c}$.
We then obtain an estimate for the effective unit of time as
\begin{equation}
\tau_\text{diffusive}= \frac{3\pi\eta \sigma_\text{L}^3}{k_\text{B}T} \frac{\langle\Delta x_\text{c}^2\rangle_\text{c}/\sigma_\text{L}^2}{\langle\Delta t_\text{c}\rangle_\text{c}/\tau}.
\end{equation}
If we use the viscosity of water $\eta=\SI{0.94}{\milli\pascal\cdot\second}$, then $3\pi\eta \sigma_\text{L}^3/k_\text{B}T = \SI{2}{\second}$.
From the analysis of collisions in the EDMD simulation shown in Fig.~3, we obtain in the fluid $\langle\Delta x_\text{c}^2\rangle_\text{c}= 10^{-3} \sigma_\text{L}^2$, $\langle\Delta t_\text{c}\rangle_\text{c} = 0.01\tau$ and 
in the fluid-solid coexistence $\langle\Delta x_\text{c}^2\rangle_\text{c}= 2 \times 10^{-3} \sigma_\text{L}^2$, $\langle\Delta t_\text{c}\rangle_\text{c} = 0.02\tau$.
The unit of time is thus estimated in both cases to be
\begin{equation}
\tau_\text{diffusive} = \SI{200}{\milli\second}.
\end{equation}
For comparison, the (much more incorrect) assumption of ballistic motion applied to polystyrene colloids gives a significantly smaller unit of time,
\begin{equation}
\tau_\text{ballistic}=\sqrt{\frac{\rho\pi\sigma_\text{L}^5}{6k_\text{B}T}}=\SI{0.4}{\milli\second}.
\end{equation}
Here, we used the density of polystyrene $\rho=\SI{1060}{\kilogram/\meter^3}$.
The crude estimate outlined above suggests that diffusion of the colloidal particles in solution slows down crystallization kinetics by up to three orders of magnitude.


%


\begin{thebibliography}{59}%
\makeatletter
\providecommand \@ifxundefined [1]{%
 \@ifx{#1\undefined}
}%
\providecommand \@ifnum [1]{%
 \ifnum #1\expandafter \@firstoftwo
 \else \expandafter \@secondoftwo
 \fi
}%
\providecommand \@ifx [1]{%
 \ifx #1\expandafter \@firstoftwo
 \else \expandafter \@secondoftwo
 \fi
}%
\providecommand \natexlab [1]{#1}%
\providecommand \enquote  [1]{``#1''}%
\providecommand \bibnamefont  [1]{#1}%
\providecommand \bibfnamefont [1]{#1}%
\providecommand \citenamefont [1]{#1}%
\providecommand \href@noop [0]{\@secondoftwo}%
\providecommand \href [0]{\begingroup \@sanitize@url \@href}%
\providecommand \@href[1]{\@@startlink{#1}\@@href}%
\providecommand \@@href[1]{\endgroup#1\@@endlink}%
\providecommand \@sanitize@url [0]{\catcode `\\12\catcode `\$12\catcode
  `\&12\catcode `\#12\catcode `\^12\catcode `\_12\catcode `\%12\relax}%
\providecommand \@@startlink[1]{}%
\providecommand \@@endlink[0]{}%
\providecommand \url  [0]{\begingroup\@sanitize@url \@url }%
\providecommand \@url [1]{\endgroup\@href {#1}{\urlprefix }}%
\providecommand \urlprefix  [0]{URL }%
\providecommand \Eprint [0]{\href }%
\providecommand \doibase [0]{http://dx.doi.org/}%
\providecommand \selectlanguage [0]{\@gobble}%
\providecommand \bibinfo  [0]{\@secondoftwo}%
\providecommand \bibfield  [0]{\@secondoftwo}%
\providecommand \translation [1]{[#1]}%
\providecommand \BibitemOpen [0]{}%
\providecommand \bibitemStop [0]{}%
\providecommand \bibitemNoStop [0]{.\EOS\space}%
\providecommand \EOS [0]{\spacefactor3000\relax}%
\providecommand \BibitemShut  [1]{\csname bibitem#1\endcsname}%
\let\auto@bib@innerbib\@empty
\bibitem [{\citenamefont {Alder}\ and\ \citenamefont
  {Wainwright}(1957)}]{Alder1957a}%
  \BibitemOpen
  \bibfield  {author} {\bibinfo {author} {\bibfnamefont {B.~J.}\ \bibnamefont
  {Alder}}\ and\ \bibinfo {author} {\bibfnamefont {T.~E.}\ \bibnamefont
  {Wainwright}},\ }\bibfield  {title} {\enquote {\bibinfo {title} {{Phase
  Transition for a Hard Sphere System}},}\ }\href {\doibase 10.1063/1.1743957}
  {\bibfield  {journal} {\bibinfo  {journal} {J. Chem. Phys.}\ }\textbf
  {\bibinfo {volume} {27}},\ \bibinfo {pages} {1208--1209} (\bibinfo {year}
  {1957})}\BibitemShut {NoStop}%
\bibitem [{\citenamefont {Wood}\ and\ \citenamefont
  {Jacobson}(1957)}]{Wood1957}%
  \BibitemOpen
  \bibfield  {author} {\bibinfo {author} {\bibfnamefont {W.~W.}\ \bibnamefont
  {Wood}}\ and\ \bibinfo {author} {\bibfnamefont {J.~D.}\ \bibnamefont
  {Jacobson}},\ }\bibfield  {title} {\enquote {\bibinfo {title} {{Preliminary
  Results from a Recalculation of the Monte Carlo Equation of State of Hard
  Spheres}},}\ }\href {\doibase 10.1063/1.1743956} {\bibfield  {journal}
  {\bibinfo  {journal} {J. Chem. Phys.}\ }\textbf {\bibinfo {volume} {27}},\
  \bibinfo {pages} {1207--1208} (\bibinfo {year} {1957})}\BibitemShut {NoStop}%
\bibitem [{\citenamefont {Alder}\ and\ \citenamefont
  {Wainwright}(1960)}]{Alder1960a}%
  \BibitemOpen
  \bibfield  {author} {\bibinfo {author} {\bibfnamefont {B.~J.}\ \bibnamefont
  {Alder}}\ and\ \bibinfo {author} {\bibfnamefont {T.~E.}\ \bibnamefont
  {Wainwright}},\ }\bibfield  {title} {\enquote {\bibinfo {title} {{Studies in
  molecular dynamics. II. Behavior of a small number of elastic spheres}},}\
  }\href {\doibase 10.1063/1.1731425} {\bibfield  {journal} {\bibinfo
  {journal} {J. Chem. Phys.}\ }\textbf {\bibinfo {volume} {33}},\ \bibinfo
  {pages} {1439--1451} (\bibinfo {year} {1960})}\BibitemShut {NoStop}%
\bibitem [{\citenamefont {Pusey}\ and\ \citenamefont {van
  Megen}(1986)}]{Pusey1986}%
  \BibitemOpen
  \bibfield  {author} {\bibinfo {author} {\bibfnamefont {P.~N.}\ \bibnamefont
  {Pusey}}\ and\ \bibinfo {author} {\bibfnamefont {W.}~\bibnamefont {van
  Megen}},\ }\bibfield  {title} {\enquote {\bibinfo {title} {{Phase behaviour
  of concentrated suspensions of nearly hard colloidal spheres}},}\ }\href
  {\doibase 10.1038/320340a0} {\bibfield  {journal} {\bibinfo  {journal}
  {Nature}\ }\textbf {\bibinfo {volume} {320}},\ \bibinfo {pages} {340--342}
  (\bibinfo {year} {1986})}\BibitemShut {NoStop}%
\bibitem [{\citenamefont {Sanders}\ and\ \citenamefont
  {Murray}(1978)}]{Sanders1978}%
  \BibitemOpen
  \bibfield  {author} {\bibinfo {author} {\bibfnamefont {J.~V.}\ \bibnamefont
  {Sanders}}\ and\ \bibinfo {author} {\bibfnamefont {M.~J.}\ \bibnamefont
  {Murray}},\ }\bibfield  {title} {\enquote {\bibinfo {title} {{Ordered
  arrangements of spheres of two different sizes in opal}},}\ }\href {\doibase
  10.1038/275201a0} {\bibfield  {journal} {\bibinfo  {journal} {Nature}\
  }\textbf {\bibinfo {volume} {275}},\ \bibinfo {pages} {201--203} (\bibinfo
  {year} {1978})}\BibitemShut {NoStop}%
\bibitem [{\citenamefont {Sanders}(1980)}]{Sanders1980}%
  \BibitemOpen
  \bibfield  {author} {\bibinfo {author} {\bibfnamefont {J.~V.}\ \bibnamefont
  {Sanders}},\ }\bibfield  {title} {\enquote {\bibinfo {title} {{Close-packed
  structures of spheres of two different sizes I. Observations on natural
  opal}},}\ }\href {\doibase 10.1080/01418618008239379} {\bibfield  {journal}
  {\bibinfo  {journal} {Philos. Mag. A}\ }\textbf {\bibinfo {volume} {42}},\
  \bibinfo {pages} {705--720} (\bibinfo {year} {1980})}\BibitemShut {NoStop}%
\bibitem [{\citenamefont {Murray}\ and\ \citenamefont
  {Sanders}(1980)}]{Murray1980}%
  \BibitemOpen
  \bibfield  {author} {\bibinfo {author} {\bibfnamefont {M.~J.}\ \bibnamefont
  {Murray}}\ and\ \bibinfo {author} {\bibfnamefont {J.~V.}\ \bibnamefont
  {Sanders}},\ }\bibfield  {title} {\enquote {\bibinfo {title} {{Close-packed
  structures of spheres of two different sizes II. The packing densities of
  likely arrangements}},}\ }\href {\doibase 10.1080/01418618008239380}
  {\bibfield  {journal} {\bibinfo  {journal} {Philos. Mag. A}\ }\textbf
  {\bibinfo {volume} {42}},\ \bibinfo {pages} {721--740} (\bibinfo {year}
  {1980})}\BibitemShut {NoStop}%
\bibitem [{\citenamefont {Bartlett}\ \emph {et~al.}(1990)\citenamefont
  {Bartlett}, \citenamefont {Ottewill},\ and\ \citenamefont
  {Pusey}}]{Bartlett1990}%
  \BibitemOpen
  \bibfield  {author} {\bibinfo {author} {\bibfnamefont {P.}~\bibnamefont
  {Bartlett}}, \bibinfo {author} {\bibfnamefont {R.~H.}\ \bibnamefont
  {Ottewill}}, \ and\ \bibinfo {author} {\bibfnamefont {P.}~\bibnamefont
  {Pusey}},\ }\bibfield  {title} {\enquote {\bibinfo {title} {{Freezing of
  binary mixtures of colloidal hard spheres}},}\ }\href {\doibase
  https://doi.org/10.1063/1.459142} {\bibfield  {journal} {\bibinfo  {journal}
  {J. Chem. Phys.}\ }\textbf {\bibinfo {volume} {93}},\ \bibinfo {pages}
  {1299--1312} (\bibinfo {year} {1990})}\BibitemShut {NoStop}%
\bibitem [{\citenamefont {Bartlett}\ \emph {et~al.}(1992)\citenamefont
  {Bartlett}, \citenamefont {Ottewill},\ and\ \citenamefont
  {Pusey}}]{Bartlett1992}%
  \BibitemOpen
  \bibfield  {author} {\bibinfo {author} {\bibfnamefont {P.}~\bibnamefont
  {Bartlett}}, \bibinfo {author} {\bibfnamefont {R.~H.}\ \bibnamefont
  {Ottewill}}, \ and\ \bibinfo {author} {\bibfnamefont {P.~N.}\ \bibnamefont
  {Pusey}},\ }\bibfield  {title} {\enquote {\bibinfo {title} {{Superlattice
  formation in binary mixtures of hard-sphere colloids}},}\ }\href {\doibase
  10.1103/PhysRevLett.68.3801} {\bibfield  {journal} {\bibinfo  {journal}
  {Phys. Rev. Lett.}\ }\textbf {\bibinfo {volume} {68}},\ \bibinfo {pages}
  {3801--3804} (\bibinfo {year} {1992})}\BibitemShut {NoStop}%
\bibitem [{\citenamefont {Hunt}\ \emph {et~al.}(2000)\citenamefont {Hunt},
  \citenamefont {Jardine},\ and\ \citenamefont {Bartlett}}]{Hunt2000}%
  \BibitemOpen
  \bibfield  {author} {\bibinfo {author} {\bibfnamefont {N.}~\bibnamefont
  {Hunt}}, \bibinfo {author} {\bibfnamefont {R.}~\bibnamefont {Jardine}}, \
  and\ \bibinfo {author} {\bibfnamefont {P.}~\bibnamefont {Bartlett}},\
  }\bibfield  {title} {\enquote {\bibinfo {title} {{Superlattice formation in
  binary mixtures of hard-sphere colloids}},}\ }\href {\doibase
  https://doi.org/10.1103/PhysRevE.62.900} {\bibfield  {journal} {\bibinfo
  {journal} {Phys. Rev. E}\ }\textbf {\bibinfo {volume} {62}},\ \bibinfo
  {pages} {900--913} (\bibinfo {year} {2000})}\BibitemShut {NoStop}%
\bibitem [{\citenamefont {Schofield}\ \emph {et~al.}(2005)\citenamefont
  {Schofield}, \citenamefont {Pusey},\ and\ \citenamefont
  {Radcliffe}}]{Schofield2005}%
  \BibitemOpen
  \bibfield  {author} {\bibinfo {author} {\bibfnamefont {A.~B.}\ \bibnamefont
  {Schofield}}, \bibinfo {author} {\bibfnamefont {P.~N.}\ \bibnamefont
  {Pusey}}, \ and\ \bibinfo {author} {\bibfnamefont {P.}~\bibnamefont
  {Radcliffe}},\ }\bibfield  {title} {\enquote {\bibinfo {title} {{Stability of
  the binary colloidal crystals AB2 and AB13}},}\ }\href {\doibase
  10.1103/PhysRevE.72.031407} {\bibfield  {journal} {\bibinfo  {journal} {Phys.
  Rev. E}\ }\textbf {\bibinfo {volume} {72}},\ \bibinfo {pages} {031407}
  (\bibinfo {year} {2005})}\BibitemShut {NoStop}%
\bibitem [{\citenamefont {Schaertl}\ \emph {et~al.}(2018)\citenamefont
  {Schaertl}, \citenamefont {Botin}, \citenamefont {Palberg},\ and\
  \citenamefont {Bartsch}}]{Schaertl2018}%
  \BibitemOpen
  \bibfield  {author} {\bibinfo {author} {\bibfnamefont {N.}~\bibnamefont
  {Schaertl}}, \bibinfo {author} {\bibfnamefont {D.}~\bibnamefont {Botin}},
  \bibinfo {author} {\bibfnamefont {T.}~\bibnamefont {Palberg}}, \ and\
  \bibinfo {author} {\bibfnamefont {E.}~\bibnamefont {Bartsch}},\ }\bibfield
  {title} {\enquote {\bibinfo {title} {{Formation of Laves phases in buoyancy
  matched hard sphere suspensions}},}\ }\href {\doibase 10.1039/C7SM02348K}
  {\bibfield  {journal} {\bibinfo  {journal} {Soft Matter}\ }\textbf {\bibinfo
  {volume} {14}},\ \bibinfo {pages} {5130--5139} (\bibinfo {year}
  {2018})}\BibitemShut {NoStop}%
\bibitem [{\citenamefont {Eldridge}\ \emph
  {et~al.}(1993{\natexlab{a}})\citenamefont {Eldridge}, \citenamefont
  {Madden},\ and\ \citenamefont {Frenkel}}]{Eldridge1993a}%
  \BibitemOpen
  \bibfield  {author} {\bibinfo {author} {\bibfnamefont {M.~D.}\ \bibnamefont
  {Eldridge}}, \bibinfo {author} {\bibfnamefont {P.~A.}\ \bibnamefont
  {Madden}}, \ and\ \bibinfo {author} {\bibfnamefont {D.}~\bibnamefont
  {Frenkel}},\ }\bibfield  {title} {\enquote {\bibinfo {title} {{A computer
  simulation investigation into the stability of the AB2 superlattice in a
  binary hard sphere system}},}\ }\href {\doibase 10.1080/00268979300102811}
  {\bibfield  {journal} {\bibinfo  {journal} {Mol. Phys.}\ }\textbf {\bibinfo
  {volume} {80}},\ \bibinfo {pages} {987--995} (\bibinfo {year}
  {1993}{\natexlab{a}})}\BibitemShut {NoStop}%
\bibitem [{\citenamefont {Eldridge}\ \emph
  {et~al.}(1993{\natexlab{b}})\citenamefont {Eldridge}, \citenamefont
  {Madden},\ and\ \citenamefont {Frenkel}}]{Eldridge1993}%
  \BibitemOpen
  \bibfield  {author} {\bibinfo {author} {\bibfnamefont {M.~D.}\ \bibnamefont
  {Eldridge}}, \bibinfo {author} {\bibfnamefont {P.~A.}\ \bibnamefont
  {Madden}}, \ and\ \bibinfo {author} {\bibfnamefont {D.}~\bibnamefont
  {Frenkel}},\ }\bibfield  {title} {\enquote {\bibinfo {title} {{Entropy-driven
  formation of a superlattice in a hard-sphere binary mixture}},}\ }\href
  {\doibase 10.1038/365035a0} {\bibfield  {journal} {\bibinfo  {journal}
  {Nature}\ }\textbf {\bibinfo {volume} {365}},\ \bibinfo {pages} {35--37}
  (\bibinfo {year} {1993}{\natexlab{b}})}\BibitemShut {NoStop}%
\bibitem [{\citenamefont {Eldridge}\ \emph
  {et~al.}(1993{\natexlab{c}})\citenamefont {Eldridge}, \citenamefont
  {Madden},\ and\ \citenamefont {Frenkel}}]{Eldridge1993b}%
  \BibitemOpen
  \bibfield  {author} {\bibinfo {author} {\bibfnamefont {M.}~\bibnamefont
  {Eldridge}}, \bibinfo {author} {\bibfnamefont {P.}~\bibnamefont {Madden}}, \
  and\ \bibinfo {author} {\bibfnamefont {D.}~\bibnamefont {Frenkel}},\
  }\bibfield  {title} {\enquote {\bibinfo {title} {{The stability of the AB13
  crystal in a binary hard sphere system}},}\ }\href {\doibase
  10.1080/00268979300101101} {\bibfield  {journal} {\bibinfo  {journal} {Mol.
  Phys.}\ }\textbf {\bibinfo {volume} {79}},\ \bibinfo {pages} {105--120}
  (\bibinfo {year} {1993}{\natexlab{c}})}\BibitemShut {NoStop}%
\bibitem [{\citenamefont {Cottin}\ and\ \citenamefont
  {Monson}(1993)}]{Cottin1993}%
  \BibitemOpen
  \bibfield  {author} {\bibinfo {author} {\bibfnamefont {X.}~\bibnamefont
  {Cottin}}\ and\ \bibinfo {author} {\bibfnamefont {P.~A.}\ \bibnamefont
  {Monson}},\ }\bibfield  {title} {\enquote {\bibinfo {title} {{A cell theory
  for solid solutions: Application to hard sphere mixtures}},}\ }\href
  {\doibase 10.1063/1.465560} {\bibfield  {journal} {\bibinfo  {journal} {J.
  Chem. Phys.}\ }\textbf {\bibinfo {volume} {99}},\ \bibinfo {pages}
  {8914--8921} (\bibinfo {year} {1993})}\BibitemShut {NoStop}%
\bibitem [{\citenamefont {Cottin}\ and\ \citenamefont
  {Monson}(1995)}]{Cottin1995}%
  \BibitemOpen
  \bibfield  {author} {\bibinfo {author} {\bibfnamefont {X.}~\bibnamefont
  {Cottin}}\ and\ \bibinfo {author} {\bibfnamefont {P.~A.}\ \bibnamefont
  {Monson}},\ }\bibfield  {title} {\enquote {\bibinfo {title}
  {{Substitutionally ordered solid solutions of hard spheres}},}\ }\href
  {\doibase 10.1063/1.469209} {\bibfield  {journal} {\bibinfo  {journal} {J.
  Chem. Phys.}\ }\textbf {\bibinfo {volume} {102}},\ \bibinfo {pages}
  {3354--3360} (\bibinfo {year} {1995})}\BibitemShut {NoStop}%
\bibitem [{\citenamefont {Trizac}\ \emph {et~al.}(1997)\citenamefont {Trizac},
  \citenamefont {Eldridge},\ and\ \citenamefont {Madden}}]{TRIZAC1997}%
  \BibitemOpen
  \bibfield  {author} {\bibinfo {author} {\bibfnamefont {E.}~\bibnamefont
  {Trizac}}, \bibinfo {author} {\bibfnamefont {M.~D.}\ \bibnamefont
  {Eldridge}}, \ and\ \bibinfo {author} {\bibfnamefont {P.~A.}\ \bibnamefont
  {Madden}},\ }\bibfield  {title} {\enquote {\bibinfo {title} {{Stability of
  the AB crystal for asymmetric binary hard sphere mixtures}},}\ }\href
  {\doibase 10.1080/002689797172408} {\bibfield  {journal} {\bibinfo  {journal}
  {Mol. Phys.}\ }\textbf {\bibinfo {volume} {90}},\ \bibinfo {pages} {675--678}
  (\bibinfo {year} {1997})}\BibitemShut {NoStop}%
\bibitem [{\citenamefont {Hynninen}\ \emph {et~al.}(2009)\citenamefont
  {Hynninen}, \citenamefont {Filion},\ and\ \citenamefont
  {Dijkstra}}]{Hynninen2009}%
  \BibitemOpen
  \bibfield  {author} {\bibinfo {author} {\bibfnamefont {A.~P.}\ \bibnamefont
  {Hynninen}}, \bibinfo {author} {\bibfnamefont {L.}~\bibnamefont {Filion}}, \
  and\ \bibinfo {author} {\bibfnamefont {M.}~\bibnamefont {Dijkstra}},\
  }\bibfield  {title} {\enquote {\bibinfo {title} {{Stability of LS and LS2
  crystal structures in binary mixtures of hard and charged spheres}},}\ }\href
  {\doibase 10.1063/1.3182724} {\bibfield  {journal} {\bibinfo  {journal} {J.
  Chem. Phys.}\ }\textbf {\bibinfo {volume} {131}},\ \bibinfo {pages} {064902}
  (\bibinfo {year} {2009})}\BibitemShut {NoStop}%
\bibitem [{\citenamefont {Dijkstra}(2014)}]{Dijkstra2014}%
  \BibitemOpen
  \bibfield  {author} {\bibinfo {author} {\bibfnamefont {M.}~\bibnamefont
  {Dijkstra}},\ }\bibfield  {title} {\enquote {\bibinfo {title}
  {{Entropy-Driven Phase Transitions in Colloids: From spheres to anisotropic
  particles}},}\ }in\ \href {\doibase 10.1002/9781118949702.ch2} {\emph
  {\bibinfo {booktitle} {Adv. Chem. Phys.}}}\ (\bibinfo {year} {2014})\ pp.\
  \bibinfo {pages} {35--71}\BibitemShut {NoStop}%
\bibitem [{\citenamefont {O'Toole}\ and\ \citenamefont
  {Hudson}(2011)}]{RN5848}%
  \BibitemOpen
  \bibfield  {author} {\bibinfo {author} {\bibfnamefont {P.~I.}\ \bibnamefont
  {O'Toole}}\ and\ \bibinfo {author} {\bibfnamefont {T.~S.}\ \bibnamefont
  {Hudson}},\ }\bibfield  {title} {\enquote {\bibinfo {title} {{New
  High-Density Packings of Similarly Sized Binary Spheres}},}\ }\href {\doibase
  10.1021/jp206115p} {\bibfield  {journal} {\bibinfo  {journal} {J. Phys. Chem.
  C}\ }\textbf {\bibinfo {volume} {115}},\ \bibinfo {pages} {19037--19040}
  (\bibinfo {year} {2011})}\BibitemShut {NoStop}%
\bibitem [{\citenamefont {Filion}\ and\ \citenamefont
  {Dijkstra}(2009)}]{Filion2009}%
  \BibitemOpen
  \bibfield  {author} {\bibinfo {author} {\bibfnamefont {L.}~\bibnamefont
  {Filion}}\ and\ \bibinfo {author} {\bibfnamefont {M.}~\bibnamefont
  {Dijkstra}},\ }\bibfield  {title} {\enquote {\bibinfo {title} {{Prediction of
  binary hard-sphere crystal structures}},}\ }\href {\doibase
  10.1103/PhysRevE.79.046714} {\bibfield  {journal} {\bibinfo  {journal} {Phys.
  Rev. E}\ }\textbf {\bibinfo {volume} {79}},\ \bibinfo {pages} {046714}
  (\bibinfo {year} {2009})}\BibitemShut {NoStop}%
\bibitem [{\citenamefont {Hopkins}\ \emph {et~al.}(2012)\citenamefont
  {Hopkins}, \citenamefont {Stillinger},\ and\ \citenamefont
  {Torquato}}]{Hopkins2012}%
  \BibitemOpen
  \bibfield  {author} {\bibinfo {author} {\bibfnamefont {A.~B.}\ \bibnamefont
  {Hopkins}}, \bibinfo {author} {\bibfnamefont {F.~H.}\ \bibnamefont
  {Stillinger}}, \ and\ \bibinfo {author} {\bibfnamefont {S.}~\bibnamefont
  {Torquato}},\ }\bibfield  {title} {\enquote {\bibinfo {title} {{Densest
  Binary Sphere Packings}},}\ }\href {\doibase 10.1103/PhysRevE.85.021130}
  {\bibfield  {journal} {\bibinfo  {journal} {Phys. Rev. E}\ }\textbf {\bibinfo
  {volume} {85}},\ \bibinfo {pages} {021130} (\bibinfo {year}
  {2012})}\BibitemShut {NoStop}%
\bibitem [{\citenamefont {Auer}\ and\ \citenamefont
  {Frenkel}(2001)}]{Auer2001}%
  \BibitemOpen
  \bibfield  {author} {\bibinfo {author} {\bibfnamefont {S.}~\bibnamefont
  {Auer}}\ and\ \bibinfo {author} {\bibfnamefont {D.}~\bibnamefont {Frenkel}},\
  }\bibfield  {title} {\enquote {\bibinfo {title} {{Prediction of absolute
  crystal-nucleation rate in hard-sphere colloids}},}\ }\href {\doibase
  10.1038/35059035} {\bibfield  {journal} {\bibinfo  {journal} {Nature}\
  }\textbf {\bibinfo {volume} {409}},\ \bibinfo {pages} {1020--1023} (\bibinfo
  {year} {2001})}\BibitemShut {NoStop}%
\bibitem [{\citenamefont {Punnathanam}\ and\ \citenamefont
  {Monson}(2006)}]{Punnathanam2006}%
  \BibitemOpen
  \bibfield  {author} {\bibinfo {author} {\bibfnamefont {S.}~\bibnamefont
  {Punnathanam}}\ and\ \bibinfo {author} {\bibfnamefont {P.~A.}\ \bibnamefont
  {Monson}},\ }\bibfield  {title} {\enquote {\bibinfo {title} {{Crystal
  nucleation in binary hard sphere mixtures: A Monte Carlo simulation
  study}},}\ }\href {\doibase 10.1063/1.2208998} {\bibfield  {journal}
  {\bibinfo  {journal} {J. Chem. Phys.}\ }\textbf {\bibinfo {volume} {125}},\
  \bibinfo {pages} {024508} (\bibinfo {year} {2006})}\BibitemShut {NoStop}%
\bibitem [{\citenamefont {Ni}\ \emph {et~al.}(2011)\citenamefont {Ni},
  \citenamefont {Smallenburg}, \citenamefont {Filion},\ and\ \citenamefont
  {Dijkstra}}]{Ni2011}%
  \BibitemOpen
  \bibfield  {author} {\bibinfo {author} {\bibfnamefont {R.}~\bibnamefont
  {Ni}}, \bibinfo {author} {\bibfnamefont {F.}~\bibnamefont {Smallenburg}},
  \bibinfo {author} {\bibfnamefont {L.}~\bibnamefont {Filion}}, \ and\ \bibinfo
  {author} {\bibfnamefont {M.}~\bibnamefont {Dijkstra}},\ }\bibfield  {title}
  {\enquote {\bibinfo {title} {{Crystal nucleation in binary hard-sphere
  mixtures: The effect of order parameter on the cluster composition}},}\
  }\href {\doibase 10.1080/00268976.2011.554333} {\bibfield  {journal}
  {\bibinfo  {journal} {Mol. Phys.}\ }\textbf {\bibinfo {volume} {109}},\
  \bibinfo {pages} {1213--1227} (\bibinfo {year} {2011})}\BibitemShut {NoStop}%
\bibitem [{\citenamefont {Vermolen}\ \emph {et~al.}(2009)\citenamefont
  {Vermolen}, \citenamefont {Kuijk}, \citenamefont {Filion}, \citenamefont
  {Hermes}, \citenamefont {Thijssen}, \citenamefont {Dijkstra},\ and\
  \citenamefont {{Van Blaaderen}}}]{Vermolen2009}%
  \BibitemOpen
  \bibfield  {author} {\bibinfo {author} {\bibfnamefont {E.~C.~M.}\
  \bibnamefont {Vermolen}}, \bibinfo {author} {\bibfnamefont {A.}~\bibnamefont
  {Kuijk}}, \bibinfo {author} {\bibfnamefont {L.~C.}\ \bibnamefont {Filion}},
  \bibinfo {author} {\bibfnamefont {M.}~\bibnamefont {Hermes}}, \bibinfo
  {author} {\bibfnamefont {J.~H.~J.}\ \bibnamefont {Thijssen}}, \bibinfo
  {author} {\bibfnamefont {M.}~\bibnamefont {Dijkstra}}, \ and\ \bibinfo
  {author} {\bibfnamefont {A.}~\bibnamefont {{Van Blaaderen}}},\ }\bibfield
  {title} {\enquote {\bibinfo {title} {{Fabrication of large binary colloidal
  crystals with a NaCl structure}},}\ }\href {\doibase 10.1073/pnas.0900605106}
  {\bibfield  {journal} {\bibinfo  {journal} {Proc. Natl. Acad. Sci. U. S. A.}\
  }\textbf {\bibinfo {volume} {106}},\ \bibinfo {pages} {16063--16067}
  (\bibinfo {year} {2009})}\BibitemShut {NoStop}%
\bibitem [{\citenamefont {Filion}\ \emph {et~al.}(2011)\citenamefont {Filion},
  \citenamefont {Hermes}, \citenamefont {Ni}, \citenamefont {Vermolen},
  \citenamefont {Kuijk}, \citenamefont {Christova}, \citenamefont
  {Stiefelhagen}, \citenamefont {Vissers}, \citenamefont {van Blaaderen},\ and\
  \citenamefont {Dijkstra}}]{Filion2011}%
  \BibitemOpen
  \bibfield  {author} {\bibinfo {author} {\bibfnamefont {L.}~\bibnamefont
  {Filion}}, \bibinfo {author} {\bibfnamefont {M.}~\bibnamefont {Hermes}},
  \bibinfo {author} {\bibfnamefont {R.}~\bibnamefont {Ni}}, \bibinfo {author}
  {\bibfnamefont {E.~C.~M.}\ \bibnamefont {Vermolen}}, \bibinfo {author}
  {\bibfnamefont {A.}~\bibnamefont {Kuijk}}, \bibinfo {author} {\bibfnamefont
  {C.~G.}\ \bibnamefont {Christova}}, \bibinfo {author} {\bibfnamefont
  {J.~C.~P.}\ \bibnamefont {Stiefelhagen}}, \bibinfo {author} {\bibfnamefont
  {T.}~\bibnamefont {Vissers}}, \bibinfo {author} {\bibfnamefont
  {A.}~\bibnamefont {van Blaaderen}}, \ and\ \bibinfo {author} {\bibfnamefont
  {M.}~\bibnamefont {Dijkstra}},\ }\bibfield  {title} {\enquote {\bibinfo
  {title} {{Self-Assembly of a Colloidal Interstitial Solid with Tunable
  Sublattice Doping}},}\ }\href {\doibase 10.1103/PhysRevLett.107.168302}
  {\bibfield  {journal} {\bibinfo  {journal} {Phys. Rev. Lett.}\ }\textbf
  {\bibinfo {volume} {107}},\ \bibinfo {pages} {168302} (\bibinfo {year}
  {2011})}\BibitemShut {NoStop}%
\bibitem [{\citenamefont {Bommineni}\ and\ \citenamefont
  {Punnathanam}(2017)}]{Bommineni2017a}%
  \BibitemOpen
  \bibfield  {author} {\bibinfo {author} {\bibfnamefont {P.~K.}\ \bibnamefont
  {Bommineni}}\ and\ \bibinfo {author} {\bibfnamefont {S.~N.}\ \bibnamefont
  {Punnathanam}},\ }\bibfield  {title} {\enquote {\bibinfo {title} {{Molecular
  simulation of homogeneous crystal nucleation of AB2 solid phase from a binary
  hard sphere mixture}},}\ }\href {\doibase 10.1063/1.4997432} {\bibfield
  {journal} {\bibinfo  {journal} {J. Chem. Phys.}\ }\textbf {\bibinfo {volume}
  {147}},\ \bibinfo {pages} {064504} (\bibinfo {year} {2017})}\BibitemShut
  {NoStop}%
\bibitem [{\citenamefont {Lindquist}\ \emph {et~al.}(2018)\citenamefont
  {Lindquist}, \citenamefont {Jadrich},\ and\ \citenamefont
  {Truskett}}]{Lindquist2018a}%
  \BibitemOpen
  \bibfield  {author} {\bibinfo {author} {\bibfnamefont {B.~A.}\ \bibnamefont
  {Lindquist}}, \bibinfo {author} {\bibfnamefont {R.~B.}\ \bibnamefont
  {Jadrich}}, \ and\ \bibinfo {author} {\bibfnamefont {T.~M.}\ \bibnamefont
  {Truskett}},\ }\bibfield  {title} {\enquote {\bibinfo {title}
  {{Communication: From close-packed to topologically close-packed: Formation
  of Laves phases in moderately polydisperse hard-sphere mixtures}},}\ }\href
  {\doibase 10.1063/1.5028279} {\bibfield  {journal} {\bibinfo  {journal} {J.
  Chem. Phys.}\ }\textbf {\bibinfo {volume} {148}},\ \bibinfo {pages} {191101}
  (\bibinfo {year} {2018})}\BibitemShut {NoStop}%
\bibitem [{\citenamefont {Bommineni}\ \emph {et~al.}(2019)\citenamefont
  {Bommineni}, \citenamefont {Varela-Rosales}, \citenamefont {Klement},\ and\
  \citenamefont {Engel}}]{Bommineni2019}%
  \BibitemOpen
  \bibfield  {author} {\bibinfo {author} {\bibfnamefont {P.~K.}\ \bibnamefont
  {Bommineni}}, \bibinfo {author} {\bibfnamefont {N.~R.}\ \bibnamefont
  {Varela-Rosales}}, \bibinfo {author} {\bibfnamefont {M.}~\bibnamefont
  {Klement}}, \ and\ \bibinfo {author} {\bibfnamefont {M.}~\bibnamefont
  {Engel}},\ }\bibfield  {title} {\enquote {\bibinfo {title} {{Complex Crystals
  from Size-Disperse Spheres}},}\ }\href {\doibase
  10.1103/PhysRevLett.122.128005} {\bibfield  {journal} {\bibinfo  {journal}
  {Phys. Rev. Lett.}\ }\textbf {\bibinfo {volume} {122}},\ \bibinfo {pages}
  {128005} (\bibinfo {year} {2019})}\BibitemShut {NoStop}%
\bibitem [{\citenamefont {Dasgupta}\ \emph {et~al.}(2020)\citenamefont
  {Dasgupta}, \citenamefont {Coli},\ and\ \citenamefont
  {Dijkstra}}]{Dasgupta2019}%
  \BibitemOpen
  \bibfield  {author} {\bibinfo {author} {\bibfnamefont {T.}~\bibnamefont
  {Dasgupta}}, \bibinfo {author} {\bibfnamefont {G.~M.}\ \bibnamefont {Coli}},
  \ and\ \bibinfo {author} {\bibfnamefont {M.}~\bibnamefont {Dijkstra}},\
  }\bibfield  {title} {\enquote {\bibinfo {title} {{Tuning the Glass
  Transition: Enhanced Crystallization of the Laves Phases in Nearly Hard
  Spheres}},}\ }\href {\doibase 10.1021/acsnano.9b07090} {\bibfield  {journal}
  {\bibinfo  {journal} {ACS Nano}\ }\textbf {\bibinfo {volume} {14}},\ \bibinfo
  {pages} {3957--3968} (\bibinfo {year} {2020})}\BibitemShut {NoStop}%
\bibitem [{\citenamefont {Steurer}\ and\ \citenamefont
  {Dshemuchadse}(2016)}]{Steurer2016}%
  \BibitemOpen
  \bibfield  {author} {\bibinfo {author} {\bibfnamefont {W.}~\bibnamefont
  {Steurer}}\ and\ \bibinfo {author} {\bibfnamefont {J.}~\bibnamefont
  {Dshemuchadse}},\ }\href {\doibase 10.1093/acprof:oso/9780198714552.001.0001}
  {\emph {\bibinfo {title} {{Intermetallics -- Structures, Properties, and
  Statistics}}}}\ (\bibinfo  {publisher} {Oxford University Press},\ \bibinfo
  {year} {2016})\BibitemShut {NoStop}%
\bibitem [{\citenamefont {Hynninen}\ \emph {et~al.}(2007)\citenamefont
  {Hynninen}, \citenamefont {Thijssen}, \citenamefont {Vermolen}, \citenamefont
  {Dijkstra},\ and\ \citenamefont {van Blaaderen}}]{Hynninen2007}%
  \BibitemOpen
  \bibfield  {author} {\bibinfo {author} {\bibfnamefont {A.-P.}\ \bibnamefont
  {Hynninen}}, \bibinfo {author} {\bibfnamefont {J.~H.~J.}\ \bibnamefont
  {Thijssen}}, \bibinfo {author} {\bibfnamefont {E.~C.~M.}\ \bibnamefont
  {Vermolen}}, \bibinfo {author} {\bibfnamefont {M.}~\bibnamefont {Dijkstra}},
  \ and\ \bibinfo {author} {\bibfnamefont {A.}~\bibnamefont {van Blaaderen}},\
  }\bibfield  {title} {\enquote {\bibinfo {title} {{Self-Assembly Route for
  Photonic Crystals with a Bandgap in the Visible Region}},}\ }\href {\doibase
  10.1038/nmat1841} {\bibfield  {journal} {\bibinfo  {journal} {Nat. Mater.}\
  }\textbf {\bibinfo {volume} {6}},\ \bibinfo {pages} {202--205} (\bibinfo
  {year} {2007})}\BibitemShut {NoStop}%
\bibitem [{\citenamefont {Berthier}\ \emph {et~al.}(2016)\citenamefont
  {Berthier}, \citenamefont {Coslovich}, \citenamefont {Ninarello},\ and\
  \citenamefont {Ozawa}}]{Berthier2016}%
  \BibitemOpen
  \bibfield  {author} {\bibinfo {author} {\bibfnamefont {L.}~\bibnamefont
  {Berthier}}, \bibinfo {author} {\bibfnamefont {D.}~\bibnamefont {Coslovich}},
  \bibinfo {author} {\bibfnamefont {A.}~\bibnamefont {Ninarello}}, \ and\
  \bibinfo {author} {\bibfnamefont {M.}~\bibnamefont {Ozawa}},\ }\bibfield
  {title} {\enquote {\bibinfo {title} {{Equilibrium sampling of hard spheres up
  to the jamming density and beyond}},}\ }\href {\doibase
  10.1103/PhysRevLett.116.238002} {\bibfield  {journal} {\bibinfo  {journal}
  {Phys. Rev. Lett.}\ }\textbf {\bibinfo {volume} {116}},\ \bibinfo {pages}
  {238002} (\bibinfo {year} {2016})}\BibitemShut {NoStop}%
\bibitem [{\citenamefont {Wyart}\ and\ \citenamefont
  {Cates}(2017)}]{Wyart2017}%
  \BibitemOpen
  \bibfield  {author} {\bibinfo {author} {\bibfnamefont {M.}~\bibnamefont
  {Wyart}}\ and\ \bibinfo {author} {\bibfnamefont {M.~E.}\ \bibnamefont
  {Cates}},\ }\bibfield  {title} {\enquote {\bibinfo {title} {{Does a Growing
  Static Length Scale Control the Glass Transition?}}}\ }\href {\doibase
  10.1103/PhysRevLett.119.195501} {\bibfield  {journal} {\bibinfo  {journal}
  {Phys. Rev. Lett.}\ }\textbf {\bibinfo {volume} {119}},\ \bibinfo {pages}
  {195501} (\bibinfo {year} {2017})}\BibitemShut {NoStop}%
\bibitem [{\citenamefont {Coslovich}\ \emph {et~al.}(2018)\citenamefont
  {Coslovich}, \citenamefont {Ozawa},\ and\ \citenamefont
  {Berthier}}]{Coslovich2018}%
  \BibitemOpen
  \bibfield  {author} {\bibinfo {author} {\bibfnamefont {D.}~\bibnamefont
  {Coslovich}}, \bibinfo {author} {\bibfnamefont {M.}~\bibnamefont {Ozawa}}, \
  and\ \bibinfo {author} {\bibfnamefont {L.}~\bibnamefont {Berthier}},\
  }\bibfield  {title} {\enquote {\bibinfo {title} {{Local order and
  crystallization of dense polydisperse hard spheres}},}\ }\href {\doibase
  10.1088/1361-648X/aab0c9} {\bibfield  {journal} {\bibinfo  {journal} {J.
  Phys. Condens. Matter}\ }\textbf {\bibinfo {volume} {30}},\ \bibinfo {pages}
  {144004} (\bibinfo {year} {2018})}\BibitemShut {NoStop}%
\bibitem [{\citenamefont {Brito}\ \emph {et~al.}(2018)\citenamefont {Brito},
  \citenamefont {Lerner},\ and\ \citenamefont {Wyart}}]{Brito2018}%
  \BibitemOpen
  \bibfield  {author} {\bibinfo {author} {\bibfnamefont {C.}~\bibnamefont
  {Brito}}, \bibinfo {author} {\bibfnamefont {E.}~\bibnamefont {Lerner}}, \
  and\ \bibinfo {author} {\bibfnamefont {M.}~\bibnamefont {Wyart}},\ }\bibfield
   {title} {\enquote {\bibinfo {title} {{Theory for swap acceleration near the
  glass and jamming transitions for continuously polydisperse particles}},}\
  }\href {\doibase 10.1103/PhysRevX.8.031050} {\bibfield  {journal} {\bibinfo
  {journal} {Phys. Rev. X}\ }\textbf {\bibinfo {volume} {8}},\ \bibinfo {pages}
  {031050} (\bibinfo {year} {2018})}\BibitemShut {NoStop}%
\bibitem [{\citenamefont {Berthier}\ \emph {et~al.}(2019)\citenamefont
  {Berthier}, \citenamefont {Flenner}, \citenamefont {Fullerton}, \citenamefont
  {Scalliet},\ and\ \citenamefont {Singh}}]{Berthier2019}%
  \BibitemOpen
  \bibfield  {author} {\bibinfo {author} {\bibfnamefont {L.}~\bibnamefont
  {Berthier}}, \bibinfo {author} {\bibfnamefont {E.}~\bibnamefont {Flenner}},
  \bibinfo {author} {\bibfnamefont {C.~J.}\ \bibnamefont {Fullerton}}, \bibinfo
  {author} {\bibfnamefont {C.}~\bibnamefont {Scalliet}}, \ and\ \bibinfo
  {author} {\bibfnamefont {M.}~\bibnamefont {Singh}},\ }\bibfield  {title}
  {\enquote {\bibinfo {title} {{Efficient swap algorithms for molecular
  dynamics simulations of equilibrium supercooled liquids}},}\ }\href {\doibase
  10.1088/1742-5468/ab1910} {\bibfield  {journal} {\bibinfo  {journal} {J.
  Stat. Mech. Theory Exp.}\ }\textbf {\bibinfo {volume} {2019}},\ \bibinfo
  {pages} {064004} (\bibinfo {year} {2019})}\BibitemShut {NoStop}%
\bibitem [{\citenamefont {Mihalkovi{\v{c}}}\ and\ \citenamefont
  {Widom}(2020)}]{Mihalkovic2019}%
  \BibitemOpen
  \bibfield  {author} {\bibinfo {author} {\bibfnamefont {M.}~\bibnamefont
  {Mihalkovi{\v{c}}}}\ and\ \bibinfo {author} {\bibfnamefont {M.}~\bibnamefont
  {Widom}},\ }\bibfield  {title} {\enquote {\bibinfo {title} {{Spontaneous
  formation of thermodynamically stable Al-Cu-Fe icosahedral quasicrystal from
  realistic atomistic simulations}},}\ }\href {\doibase
  10.1103/PhysRevResearch.2.013196} {\bibfield  {journal} {\bibinfo  {journal}
  {Phys. Rev. Res.}\ }\textbf {\bibinfo {volume} {2}},\ \bibinfo {pages}
  {013196} (\bibinfo {year} {2020})}\BibitemShut {NoStop}%
\bibitem [{Sup()}]{SuppMat}%
  \BibitemOpen
  \href@noop {} {\enquote {\bibinfo {title} {{See Supplemental Material at
  http://link.aps.org/ supplemental/10.1103/PhysRevLett.xxx.xxxxxx for
  additional text on simulation methods, detection of crystalline clusters,
  mapping of simulations to experiments, and a movie of Laves phase growth.}}}\
  }\BibitemShut {NoStop}%
\bibitem [{\citenamefont {Steinhardt}\ \emph {et~al.}(1983)\citenamefont
  {Steinhardt}, \citenamefont {Nelson},\ and\ \citenamefont
  {Ronchetti}}]{Steinhardt1983}%
  \BibitemOpen
  \bibfield  {author} {\bibinfo {author} {\bibfnamefont {P.~J.}\ \bibnamefont
  {Steinhardt}}, \bibinfo {author} {\bibfnamefont {D.~R.}\ \bibnamefont
  {Nelson}}, \ and\ \bibinfo {author} {\bibfnamefont {M.}~\bibnamefont
  {Ronchetti}},\ }\bibfield  {title} {\enquote {\bibinfo {title}
  {{Bond-orientational order in liquids and glasses}},}\ }\href {\doibase
  10.1103/PhysRevB.28.784} {\bibfield  {journal} {\bibinfo  {journal} {Phys.
  Rev. B}\ }\textbf {\bibinfo {volume} {28}},\ \bibinfo {pages} {784--805}
  (\bibinfo {year} {1983})}\BibitemShut {NoStop}%
\bibitem [{\citenamefont {van Meel}\ \emph {et~al.}(2012)\citenamefont {van
  Meel}, \citenamefont {Filion}, \citenamefont {Valeriani},\ and\ \citenamefont
  {Frenkel}}]{vanMeel2012}%
  \BibitemOpen
  \bibfield  {author} {\bibinfo {author} {\bibfnamefont {J.~A.}\ \bibnamefont
  {van Meel}}, \bibinfo {author} {\bibfnamefont {L.}~\bibnamefont {Filion}},
  \bibinfo {author} {\bibfnamefont {C.}~\bibnamefont {Valeriani}}, \ and\
  \bibinfo {author} {\bibfnamefont {D.}~\bibnamefont {Frenkel}},\ }\bibfield
  {title} {\enquote {\bibinfo {title} {{A parameter-free, solid-angle based,
  nearest-neighbor algorithm}},}\ }\href {\doibase 10.1063/1.4729313}
  {\bibfield  {journal} {\bibinfo  {journal} {J. Chem. Phys.}\ }\textbf
  {\bibinfo {volume} {136}},\ \bibinfo {pages} {234107} (\bibinfo {year}
  {2012})}\BibitemShut {NoStop}%
\bibitem [{\citenamefont {Engel}\ \emph {et~al.}(2015)\citenamefont {Engel},
  \citenamefont {Damasceno}, \citenamefont {Phillips},\ and\ \citenamefont
  {Glotzer}}]{Engel2015}%
  \BibitemOpen
  \bibfield  {author} {\bibinfo {author} {\bibfnamefont {M.}~\bibnamefont
  {Engel}}, \bibinfo {author} {\bibfnamefont {P.~F.}\ \bibnamefont
  {Damasceno}}, \bibinfo {author} {\bibfnamefont {C.~L.}\ \bibnamefont
  {Phillips}}, \ and\ \bibinfo {author} {\bibfnamefont {S.~C.}\ \bibnamefont
  {Glotzer}},\ }\bibfield  {title} {\enquote {\bibinfo {title} {{Computational
  self-assembly of a one-component icosahedral quasicrystal}},}\ }\href
  {\doibase 10.1038/nmat4152} {\bibfield  {journal} {\bibinfo  {journal} {Nat.
  Mater.}\ }\textbf {\bibinfo {volume} {14}},\ \bibinfo {pages} {109--116}
  (\bibinfo {year} {2015})}\BibitemShut {NoStop}%
\bibitem [{\citenamefont {Klement}\ and\ \citenamefont
  {Engel}(2019)}]{Klement2019}%
  \BibitemOpen
  \bibfield  {author} {\bibinfo {author} {\bibfnamefont {M.}~\bibnamefont
  {Klement}}\ and\ \bibinfo {author} {\bibfnamefont {M.}~\bibnamefont
  {Engel}},\ }\bibfield  {title} {\enquote {\bibinfo {title} {{Efficient
  equilibration of hard spheres with Newtonian event chains}},}\ }\href
  {\doibase 10.1063/1.5090882} {\bibfield  {journal} {\bibinfo  {journal} {J.
  Chem. Phys.}\ }\textbf {\bibinfo {volume} {150}},\ \bibinfo {pages} {174108}
  (\bibinfo {year} {2019})}\BibitemShut {NoStop}%
\bibitem [{\citenamefont {Tateno}\ \emph {et~al.}(2019)\citenamefont {Tateno},
  \citenamefont {Yanagishima}, \citenamefont {Russo},\ and\ \citenamefont
  {Tanaka}}]{Tateno2019}%
  \BibitemOpen
  \bibfield  {author} {\bibinfo {author} {\bibfnamefont {M.}~\bibnamefont
  {Tateno}}, \bibinfo {author} {\bibfnamefont {T.}~\bibnamefont {Yanagishima}},
  \bibinfo {author} {\bibfnamefont {J.}~\bibnamefont {Russo}}, \ and\ \bibinfo
  {author} {\bibfnamefont {H.}~\bibnamefont {Tanaka}},\ }\bibfield  {title}
  {\enquote {\bibinfo {title} {{Influence of Hydrodynamic Interactions on
  Colloidal Crystallization}},}\ }\href {\doibase
  10.1103/PhysRevLett.123.258002} {\bibfield  {journal} {\bibinfo  {journal}
  {Phys. Rev. Lett.}\ }\textbf {\bibinfo {volume} {123}},\ \bibinfo {pages}
  {258002} (\bibinfo {year} {2019})}\BibitemShut {NoStop}%
\bibitem [{\citenamefont {Hachisu}\ and\ \citenamefont
  {Yoshimura}(1980)}]{Hachisu1980}%
  \BibitemOpen
  \bibfield  {author} {\bibinfo {author} {\bibfnamefont {S.}~\bibnamefont
  {Hachisu}}\ and\ \bibinfo {author} {\bibfnamefont {S.}~\bibnamefont
  {Yoshimura}},\ }\bibfield  {title} {\enquote {\bibinfo {title} {{Optical
  demonstration of crystalline superstructures in binary mixtures of latex
  globules}},}\ }\href {\doibase 10.1038/283188a0} {\bibfield  {journal}
  {\bibinfo  {journal} {Nature}\ }\textbf {\bibinfo {volume} {283}},\ \bibinfo
  {pages} {188--189} (\bibinfo {year} {1980})}\BibitemShut {NoStop}%
\bibitem [{\citenamefont {Bartlett}\ and\ \citenamefont
  {Campbell}(2005)}]{Bartlett2005}%
  \BibitemOpen
  \bibfield  {author} {\bibinfo {author} {\bibfnamefont {P.}~\bibnamefont
  {Bartlett}}\ and\ \bibinfo {author} {\bibfnamefont {A.~I.}\ \bibnamefont
  {Campbell}},\ }\bibfield  {title} {\enquote {\bibinfo {title}
  {{Three-dimensional binary superlattices of oppositely charged colloids}},}\
  }\href {\doibase 10.1103/PhysRevLett.95.128302} {\bibfield  {journal}
  {\bibinfo  {journal} {Phys. Rev. Lett.}\ }\textbf {\bibinfo {volume} {95}},\
  \bibinfo {pages} {128302} (\bibinfo {year} {2005})}\BibitemShut {NoStop}%
\bibitem [{\citenamefont {Leunissen}\ \emph {et~al.}(2005)\citenamefont
  {Leunissen}, \citenamefont {Christova}, \citenamefont {Hynninen},
  \citenamefont {Royall}, \citenamefont {Campbell}, \citenamefont {Imhof},
  \citenamefont {Dijkstra}, \citenamefont {van Roij},\ and\ \citenamefont {van
  Blaaderen}}]{Leunissen2005}%
  \BibitemOpen
  \bibfield  {author} {\bibinfo {author} {\bibfnamefont {M.~E.}\ \bibnamefont
  {Leunissen}}, \bibinfo {author} {\bibfnamefont {C.~G.}\ \bibnamefont
  {Christova}}, \bibinfo {author} {\bibfnamefont {A.~P.}\ \bibnamefont
  {Hynninen}}, \bibinfo {author} {\bibfnamefont {C.~P.}\ \bibnamefont
  {Royall}}, \bibinfo {author} {\bibfnamefont {A.~I.}\ \bibnamefont
  {Campbell}}, \bibinfo {author} {\bibfnamefont {A.}~\bibnamefont {Imhof}},
  \bibinfo {author} {\bibfnamefont {M.}~\bibnamefont {Dijkstra}}, \bibinfo
  {author} {\bibfnamefont {R.}~\bibnamefont {van Roij}}, \ and\ \bibinfo
  {author} {\bibfnamefont {A.}~\bibnamefont {van Blaaderen}},\ }\bibfield
  {title} {\enquote {\bibinfo {title} {{Ionic colloidal crystals of oppositely
  charged particles}},}\ }\href {\doibase 10.1038/nature03946} {\bibfield
  {journal} {\bibinfo  {journal} {Nature}\ }\textbf {\bibinfo {volume} {437}},\
  \bibinfo {pages} {235--240} (\bibinfo {year} {2005})}\BibitemShut {NoStop}%
\bibitem [{\citenamefont {LaCour}\ \emph {et~al.}(2019)\citenamefont {LaCour},
  \citenamefont {Adorf}, \citenamefont {Dshemuchadse},\ and\ \citenamefont
  {Glotzer}}]{LaCour2019}%
  \BibitemOpen
  \bibfield  {author} {\bibinfo {author} {\bibfnamefont {R.~A.}\ \bibnamefont
  {LaCour}}, \bibinfo {author} {\bibfnamefont {C.~S.}\ \bibnamefont {Adorf}},
  \bibinfo {author} {\bibfnamefont {J.}~\bibnamefont {Dshemuchadse}}, \ and\
  \bibinfo {author} {\bibfnamefont {S.~C.}\ \bibnamefont {Glotzer}},\
  }\bibfield  {title} {\enquote {\bibinfo {title} {{Influence of Softness on
  the Stability of Binary Colloidal Crystals}},}\ }\href {\doibase
  10.1021/acsnano.9b04274} {\bibfield  {journal} {\bibinfo  {journal} {ACS
  Nano}\ }\textbf {\bibinfo {volume} {13}},\ \bibinfo {pages} {13829--13842}
  (\bibinfo {year} {2019})}\BibitemShut {NoStop}%
\bibitem [{\citenamefont {Shevchenko}\ \emph {et~al.}(2005)\citenamefont
  {Shevchenko}, \citenamefont {Talapin}, \citenamefont {O'Brien},\ and\
  \citenamefont {Murray}}]{Shevchenko2005}%
  \BibitemOpen
  \bibfield  {author} {\bibinfo {author} {\bibfnamefont {E.~V.}\ \bibnamefont
  {Shevchenko}}, \bibinfo {author} {\bibfnamefont {D.~V.}\ \bibnamefont
  {Talapin}}, \bibinfo {author} {\bibfnamefont {S.}~\bibnamefont {O'Brien}}, \
  and\ \bibinfo {author} {\bibfnamefont {C.~B.}\ \bibnamefont {Murray}},\
  }\bibfield  {title} {\enquote {\bibinfo {title} {{Polymorphism in AB13
  nanoparticle superlattices: An example of semiconductor-metal
  metamaterials}},}\ }\href {\doibase 10.1021/ja050510z} {\bibfield  {journal}
  {\bibinfo  {journal} {J. Am. Chem. Soc.}\ }\textbf {\bibinfo {volume}
  {127}},\ \bibinfo {pages} {8741--8747} (\bibinfo {year} {2005})}\BibitemShut
  {NoStop}%
\bibitem [{\citenamefont {Shevchenko}\ \emph {et~al.}(2006)\citenamefont
  {Shevchenko}, \citenamefont {Talapin}, \citenamefont {Kotov}, \citenamefont
  {O'Brien},\ and\ \citenamefont {Murray}}]{Shevchenko2006}%
  \BibitemOpen
  \bibfield  {author} {\bibinfo {author} {\bibfnamefont {E.~V.}\ \bibnamefont
  {Shevchenko}}, \bibinfo {author} {\bibfnamefont {D.~V.}\ \bibnamefont
  {Talapin}}, \bibinfo {author} {\bibfnamefont {N.~A.}\ \bibnamefont {Kotov}},
  \bibinfo {author} {\bibfnamefont {S.}~\bibnamefont {O'Brien}}, \ and\
  \bibinfo {author} {\bibfnamefont {C.~B.}\ \bibnamefont {Murray}},\ }\bibfield
   {title} {\enquote {\bibinfo {title} {{Structural diversity in binary
  nanoparticle superlattices}},}\ }\href {\doibase 10.1038/nature04414}
  {\bibfield  {journal} {\bibinfo  {journal} {Nature}\ }\textbf {\bibinfo
  {volume} {439}},\ \bibinfo {pages} {55--59} (\bibinfo {year}
  {2006})}\BibitemShut {NoStop}%
\bibitem [{\citenamefont {Evers}\ \emph {et~al.}(2010)\citenamefont {Evers},
  \citenamefont {Nijs}, \citenamefont {Filion}, \citenamefont {Castillo},
  \citenamefont {Dijkstra},\ and\ \citenamefont {Vanmaekelbergh}}]{Evers2010a}%
  \BibitemOpen
  \bibfield  {author} {\bibinfo {author} {\bibfnamefont {W.~H.}\ \bibnamefont
  {Evers}}, \bibinfo {author} {\bibfnamefont {B.~D.}\ \bibnamefont {Nijs}},
  \bibinfo {author} {\bibfnamefont {L.}~\bibnamefont {Filion}}, \bibinfo
  {author} {\bibfnamefont {S.}~\bibnamefont {Castillo}}, \bibinfo {author}
  {\bibfnamefont {M.}~\bibnamefont {Dijkstra}}, \ and\ \bibinfo {author}
  {\bibfnamefont {D.}~\bibnamefont {Vanmaekelbergh}},\ }\bibfield  {title}
  {\enquote {\bibinfo {title} {{Entropy-driven formation of binary
  semiconductor-nanocrystal superlattices}},}\ }\href {\doibase
  10.1021/nl102705p} {\bibfield  {journal} {\bibinfo  {journal} {Nano Lett.}\
  }\textbf {\bibinfo {volume} {10}},\ \bibinfo {pages} {4235--4241} (\bibinfo
  {year} {2010})}\BibitemShut {NoStop}%
\bibitem [{\citenamefont {Coropceanu}\ \emph {et~al.}(2019)\citenamefont
  {Coropceanu}, \citenamefont {Boles},\ and\ \citenamefont
  {Talapin}}]{Coropceanu2019}%
  \BibitemOpen
  \bibfield  {author} {\bibinfo {author} {\bibfnamefont {I.}~\bibnamefont
  {Coropceanu}}, \bibinfo {author} {\bibfnamefont {M.~A.}\ \bibnamefont
  {Boles}}, \ and\ \bibinfo {author} {\bibfnamefont {D.~V.}\ \bibnamefont
  {Talapin}},\ }\bibfield  {title} {\enquote {\bibinfo {title} {{Systematic
  Mapping of Binary Nanocrystal Superlattices: The Role of Topology in Phase
  Selection}},}\ }\href {\doibase 10.1021/jacs.8b12539} {\bibfield  {journal}
  {\bibinfo  {journal} {J. Am. Chem. Soc.}\ }\textbf {\bibinfo {volume}
  {141}},\ \bibinfo {pages} {5728--5740} (\bibinfo {year} {2019})}\BibitemShut
  {NoStop}%
\bibitem [{\citenamefont {Pedersen}\ \emph {et~al.}(2018)\citenamefont
  {Pedersen}, \citenamefont {Schr{\o}der},\ and\ \citenamefont
  {Dyre}}]{Pedersen2018}%
  \BibitemOpen
  \bibfield  {author} {\bibinfo {author} {\bibfnamefont {U.~R.}\ \bibnamefont
  {Pedersen}}, \bibinfo {author} {\bibfnamefont {T.~B.}\ \bibnamefont
  {Schr{\o}der}}, \ and\ \bibinfo {author} {\bibfnamefont {J.~C.}\ \bibnamefont
  {Dyre}},\ }\bibfield  {title} {\enquote {\bibinfo {title} {{Phase Diagram of
  Kob-Andersen-Type Binary Lennard-Jones Mixtures}},}\ }\href {\doibase
  10.1103/PhysRevLett.120.165501} {\bibfield  {journal} {\bibinfo  {journal}
  {Phys. Rev. Lett.}\ }\textbf {\bibinfo {volume} {120}},\ \bibinfo {pages}
  {165501} (\bibinfo {year} {2018})}\BibitemShut {NoStop}%
\bibitem [{\citenamefont {Ingebrigtsen}\ \emph {et~al.}(2019)\citenamefont
  {Ingebrigtsen}, \citenamefont {Dyre}, \citenamefont {Schr{\o}der},\ and\
  \citenamefont {Royall}}]{Ingebrigtsen2019}%
  \BibitemOpen
  \bibfield  {author} {\bibinfo {author} {\bibfnamefont {T.~S.}\ \bibnamefont
  {Ingebrigtsen}}, \bibinfo {author} {\bibfnamefont {J.~C.}\ \bibnamefont
  {Dyre}}, \bibinfo {author} {\bibfnamefont {T.~B.}\ \bibnamefont
  {Schr{\o}der}}, \ and\ \bibinfo {author} {\bibfnamefont {C.~P.}\ \bibnamefont
  {Royall}},\ }\bibfield  {title} {\enquote {\bibinfo {title} {{Crystallization
  Instability in Glass-Forming Mixtures}},}\ }\href {\doibase
  10.1103/PhysRevX.9.031016} {\bibfield  {journal} {\bibinfo  {journal} {Phys.
  Rev. X}\ }\textbf {\bibinfo {volume} {9}},\ \bibinfo {pages} {031016}
  (\bibinfo {year} {2019})}\BibitemShut {NoStop}%
\bibitem [{\citenamefont {Robinson}\ \emph {et~al.}(2019)\citenamefont
  {Robinson}, \citenamefont {Turci}, \citenamefont {Roth},\ and\ \citenamefont
  {Royall}}]{Robinson2019}%
  \BibitemOpen
  \bibfield  {author} {\bibinfo {author} {\bibfnamefont {J.~F.}\ \bibnamefont
  {Robinson}}, \bibinfo {author} {\bibfnamefont {F.}~\bibnamefont {Turci}},
  \bibinfo {author} {\bibfnamefont {R.}~\bibnamefont {Roth}}, \ and\ \bibinfo
  {author} {\bibfnamefont {C.~P.}\ \bibnamefont {Royall}},\ }\bibfield  {title}
  {\enquote {\bibinfo {title} {{Morphometric Approach to Many-Body Correlations
  in Hard Spheres}},}\ }\href {\doibase 10.1103/PhysRevLett.122.068004}
  {\bibfield  {journal} {\bibinfo  {journal} {Phys. Rev. Lett.}\ }\textbf
  {\bibinfo {volume} {122}},\ \bibinfo {pages} {068004} (\bibinfo {year}
  {2019})}\BibitemShut {NoStop}%
\bibitem [{\citenamefont {Mar{\'{i}}n-Aguilar}\ \emph
  {et~al.}(2020)\citenamefont {Mar{\'{i}}n-Aguilar}, \citenamefont {Wensink},
  \citenamefont {Foffi},\ and\ \citenamefont
  {Smallenburg}}]{Marin-Aguilar2019}%
  \BibitemOpen
  \bibfield  {author} {\bibinfo {author} {\bibfnamefont {S.}~\bibnamefont
  {Mar{\'{i}}n-Aguilar}}, \bibinfo {author} {\bibfnamefont {H.~H.}\
  \bibnamefont {Wensink}}, \bibinfo {author} {\bibfnamefont {G.}~\bibnamefont
  {Foffi}}, \ and\ \bibinfo {author} {\bibfnamefont {F.}~\bibnamefont
  {Smallenburg}},\ }\bibfield  {title} {\enquote {\bibinfo {title}
  {{Tetrahedrality dictates dynamics in hard sphere mixtures}},}\ }\href
  {http://arxiv.org/abs/1908.00425} {\bibfield  {journal} {\bibinfo  {journal}
  {Phys. Rev. Lett.}\ }\textbf {\bibinfo {volume} {in press}} (\bibinfo {year}
  {2020})}\BibitemShut {NoStop}%
\bibitem [{\citenamefont {Campo}\ and\ \citenamefont
  {Speck}(2020)}]{Campo2019}%
  \BibitemOpen
  \bibfield  {author} {\bibinfo {author} {\bibfnamefont {M.}~\bibnamefont
  {Campo}}\ and\ \bibinfo {author} {\bibfnamefont {T.}~\bibnamefont {Speck}},\
  }\bibfield  {title} {\enquote {\bibinfo {title} {{Dynamical coexistence in
  moderately polydisperse hard-sphere glasses}},}\ }\href {\doibase
  10.1063/1.5134842} {\bibfield  {journal} {\bibinfo  {journal} {J. Chem.
  Phys.}\ }\textbf {\bibinfo {volume} {152}},\ \bibinfo {pages} {014501}
  (\bibinfo {year} {2020})}\BibitemShut {NoStop}%
\end{thebibliography}
\end{document}